\documentclass[universe,article,accept,oneauthors,pdftex,12pt,a4paper]{mdpi}
\lastpage{\pageref*{LastPage}}
\doinum{10.3390/universe1030446}
\pubvolume{1}
\pubyear{2015}
\externaleditor{Academic Editors: Kazuharu Bamba and Sergei D. Odintsov}
\history{Received: 10 October 2015 / Accepted: 16 November 2015 / Published: 1 December 2015}

\usepackage{amsfonts}
\usepackage{amsmath}
\usepackage{amssymb}
\usepackage{mathtools}
\usepackage{mathrsfs}
\usepackage{enumerate}
\usepackage{bbm}
\usepackage{graphicx,epsfig,amsthm,color}
\usepackage{pstricks}
\usepackage{subfig}
\usepackage{graphicx}
\usepackage{units}
\usepackage{caption}
\usepackage{booktabs}
\usepackage{textcomp}
\usepackage{array}
\usepackage{cancel}

\usepackage{amssymb}
\usepackage{xcolor}
\usepackage[normalem]{ulem} 
\newcommand\hl{\bgroup\markoverwith
{\textcolor{yellow}{\rule[-.5ex]{2pt}{2.5ex}}}\ULon}
\usepackage{upgreek}

\newcolumntype{z}[1]{>{\RaggedRight\hspace{0pt}}p{#1}}
\newcolumntype{w}[1]{>{\RaggedRight\hspace{0pt}}p{#1}}
\newcolumntype{v}[1]{>{\Centering\hspace{0pt}}p{#1}}

\def\be{\begin{equation}}
\def\ee{\end{equation}}
\def\bea{\begin{eqnarray}}
\def\eea{\end{eqnarray}}

\def\be{\begin{equation}}
\def\ee{\end{equation}}
\def\bea{\begin{eqnarray}}
\def\eea{\end{eqnarray}}

\def\erp2{{\rm e}^{2\rho}}
\def\erm2{{\rm e}^{-2\rho}}
\def\er4{{\rm e}^{4\rho}}

\def\be{\begin{equation}}
\def\ee{\end{equation}}
\def\bea{\begin{eqnarray}}
\def\eea{\end{eqnarray}}

\def\m0{m_{\nu_{0,i}}}

\def\T0{T_{\nu_0}}

\renewcommand{\labelenumi}{[\arabic{enumi}]}

\newcommand{\half}{\frac{1}{2}}

\newcommand{\beqa}{\begin{eqnarray}}
\newcommand{\eeqa}{\end{eqnarray}}
\newcommand{\bpr}{\begin{problem}}
\newcommand{\epr}{\end{problem}}
\newcommand{\bcent}{\begin{center}}
\newcommand{\ecent}{\end{center}}
\newcommand{\bfig}{\begin{figure}}
\newcommand{\efig}{\end{figure}}
\newcommand{\bpc}{\begin{picture}}
\newcommand{\epc}{\end{picture}}

\renewcommand{\and}{A_{0}^{\nu ,D}(s)}

\newcommand{\bee}{\begin{equation}}

\def\beq{\begin{eqnarray}}
\def\eeq{\end{eqnarray}}

\newcommand{\bright}{\begin{flushright}}
\newcommand{\eright}{\end{flushright}}
\newcommand{\bminip}{\begin{minipage}}
\newcommand{\eminip}{\end{minipage}}

\Title{ Chameleonic Theories: A Short Review}

\Author{Andrea Zanzi}

\address[1]{%
Dipartimento di Fisica e Scienze della Terra, Universit\'a di Ferrara, Ferrara 44100, Italy; \linebreak E-Mail: andrea.zanzi@unife.it}

\corres{andrea.zanzi@unife.it}

\abstract{In the chameleon mechanism, a field (typically scalar) has a mass that depends on the matter density of the environment: the larger is the matter density, the larger is the mass of the chameleon. We briefly review some aspects of chameleonic theories. \mbox{In particular,} in a typical class of these theories, we discuss the lagrangian, the role of conformal transformations, the equation of motion and the thin-shell effect. We also discuss{ $f(R)$} theories and chameleonic quantum gravity.}


\begin{document}

\setcounter{equation}{0}
\section{Introduction}

The Standard Model (SM) of electroweak and strong interactions provides an extremely successful description of particle physics up to the (roughly) TeV scale. However, there are reasons to go beyond the SM (for an introduction see \cite{Kazakov:2006kp}). Interestingly, a plethora of scalar degrees of freedom can appear into the effective theories describing the low-energy regime of these extensions of the SM.

One remarkable example is given by string theory (see, for example, \cite{Luest:2013ll}), where gauge neutral scalar fields with perturbatively flat potential (\emph{i.e.}, moduli fields) are present. These moduli can be coupled to matter and they can control the value of the fundamental ``constants''. Therefore, in order to construct a realistic string model, the stabilization of the moduli is, typically, a crucial step. In other words, if we stabilize (\emph{i.e.}, we give a large mass to) the moduli we can obtain a model such that (1) the range of the modulus-mediated interaction is short and (2) the fundamental ``constants'' are really fixed. In a realistic scenario, the conditions (1) and (2) are certainly welcome and many efforts have been dedicated to moduli stabilization (see for example \cite{Correia:2007sv} and related references).

This review will discuss the so-called ``chameleon'' mechanism. It can be considered a stabilization mechanism, but it is very peculiar, because it is related to the large density of the local environment. \mbox{The chameleon} mechanism is also a {{ screening}}  mechanism: the deviations from standard General Relativity (GR) are suppressed locally exploiting the large ({{{Large with respect} to the cosmological value}}) density of the local environment. To date, three screening mechanisms turned out to be successful:
\begin{itemize}
\item {The chameleon mechanism \cite{Khoury:2003aq, Khoury:2003rn}, where a field (typically scalar) has a mass that depends on the matter density of the environment: the larger is the matter density, the larger is the mass of the chameleon. On cosmological distances, where the matter density is small, the chameleon is ultralight and, hence, it can play the role of a sort of quintessence field. Interestingly, in \cite{Zanzi:2010rs, Zanzi:2012du, Zanzi:2012ha, Zanzi:2012bf, Zanzi:2014twa} a chameleonic model that keeps under control the cosmological constant without fine-tuning of the parameters and including all quantum contributions has been discussed.}\vspace{-9pt}
\item {The Vainshtein mechanism \cite{Vainshtein:1972sx} takes advantage of the non-linearities sourced by the self-coupling of a scalar field.}\vspace{-9pt}
\item {The symmetron mechanism \cite{Hinterbichler:2010es, Pietroni:2005pv, Olive:2007aj} where fifth forces are screened through the restoration of a symmetry at high-density.}
\end{itemize}

Let us discuss the first mechanism. Two issues are particularly relevant. On the one hand, stronger theoretical grounds supporting chameleonic theories are welcome (many efforts have been dedicated to this topic---see for example \cite{Zanzi:2010rs, Zanzi:2012bf, Hinterbichler:2010wu, Nastase:2013ik, Brax:2011qs, Brax:2012mq, Hinterbichler:2013we}), on the other hand, experimental signatures of chameleons.

Let us further elaborate the point we mentioned last. At the moment, it is not known whether chameleon fields are present in Nature or not. The search for chameleon fields has been performed in various ways. Let us summarize some of them:

\begin{enumerate}
\renewcommand{\labelenumi}{(\arabic{enumi})}
\item{\it {In} the Laboratory}.
 In this case we mention three experiments: the E$\ddot{o}$t-Wash \mbox{experiment \cite{Adelberger:2006dh},} the CHASE \mbox{experiment \cite{Steffen:2010ze},} the {ADMX} experiment (Axion Dark Matter eXperiment) \cite{Rybka:2010ah}. The first one is designed to investigate potential deviations from the inverse-square law of gravity on length scales larger than (roughly) 50 $\upmu$m. Let us now briefly touch upon the remaining two experiments. CHASE and ADMX experiments are based on the potential presence of a coupling between the chameleon field and the $F_{\mu\nu}F^{\mu\nu}$ term of photons. This coupling, in the presence of a magnetic field, can lead to oscillations photons-chameleons that might be detected by CHASE and ADMX.
\item {\it {Astrophysically}}.
As already discussed in \cite{Hu:2013aqa}, PLANCK data put tight constraints on $\beta_1=1.043^{+0.163}_{-0.104}$ (at $95\%$ C.L.) which parametrizes the non-minimal coupling chameleon-matter (in GR we have $\beta_1=1$). PLANCK data are not the only source of astrophysical/cosmological constraints. For example, the photon-chameleon mixing mentioned above can occur inside the Sun \cite{Brax:2010xq} and this might explain the solar corona problem. The same mixing photons-chameleons can occur inside the magnetic field of the Coma cluster: in this case a chameleonic Sunyaev-Zeldovich (CSZ) effect has been predicted. In other words, the CSZ effect is a reduction of the overall photon intensity due to the conversion photon-chameleon inside the magnetic field. It is similar to the standard {SZ} effect due to the interaction photons-electrons. However, there is a difference: the standard SZ effect decreases rapidly towards the edge of the cluster where the number density of the electrons is small while, by contrast, the CSZ effect depends on the ratio of the magnetic field to the electron density and, therefore, it might be the origin of the greater than expected SZ signal detected at large radii \cite{Davis:2009vk, Davis:2010nj}. Moreover, chameleons affect the internal dynamics and stellar evolution in dwarf galaxies in regions with a sufficiently small density (see \cite{Khoury:2013yy} and references therein). For a discussion of the link between chameleon fields and helioseismology see \cite{Zanzi:2014aia}.
\item{\it {Performing} Tests of Gravity in Space}.
Interestingly, small bodies that are screened in the laboratory can be unscreened in space, where the matter density is much smaller. Hence, chameleons can produce violations of the weak equivalence principle (WEP) in orbit with $\eta=\Delta a/a >> 10^{-13}$, in conflict with the corresponding laboratory constraints. Analogously, $\cal{O}$(1) deviations from the value of $G_N$ measured on Earth are expected in the theory. These properties of chameleons lead to interesting predictions for future satellite experiments designed to test gravity in space such as MicroSCOPE ({{http://microscope.onera.fr/}}) and STE-QUEST ({{http://sci.esa.int/science-e/www/area/index.cfm?fareaid=127}}.)
\end{enumerate}

Interestingly, chameleon fields are relevant not only for phenomenological reasons, but also for theoretical ones. For example they are a useful guideline towards the construction of a quantum theory of gravitation. Indeed, in \cite{Zanzi:2014twa} a Chameleonic Equivalence Principle (CEP) has been formulated as a consequence of the chameleon mechanism. The CEP establishes an equivalence between a conformal anomaly and the quantum gravitational field. This principle is formulated in the so-called Modified Fujii's Model (MFM) which has been exploited in \cite{Zanzi:2010rs} to solve the Cosmological Constant (CC) problem. Let us further discuss this issue. In the E-frame of the MFM a chameleonic dilaton $\sigma$ is parametrizing the scale invariance of the model. Locally, in the {UV} {(ultraviolet)}, scale invariance is abundantly broken, particle masses are large, the vacuum energy is large. On the contrary, globally, in the {IR} region, particles are very light, the renormalized vacuum energy is small and scale invariance is basically restored. The non-linear nature of the chameleonic theory is crucial to keep under control the CC. The reader is referred to \cite{Zanzi:2010rs} for more details. It is common knowledge that the CC problem is really acute only in the quantum gravity regime and, hence, the CEP is a step beyond the (basically) semi-classical analysis of \cite{Zanzi:2010rs}. \mbox{Another theoretical} reason to consider chameleon fields is that they can describe (in the MFM) the collapse of the wave function in quantum mechanics \cite{Zanzi:2014twa} and the CEP is telling us that the collapse is a quantum gravity effect. For example, let us consider a diffraction experiment with electrons (forming a plane wave) scattered through a circular hole. The system is axially-symmetric. When we perform a quantum measurement, namely, when the electrons enter into the screen, there is a shift in the matter density of the environment, the chameleon jumps to another ground state and the harmonic approximation is broken during the jump. Therefore, the non-linear nature of the theory breaks the superposition principle for a short time. The expected diffraction pattern will respect the circular symmetry (\emph{i.e.}, it corresponds to a set of degenerate states), but the single electron on the screen will {{ break}} the symmetry. \mbox{Where does} this symmetry breaking come from? The chameleonic lagrangian is rotationally symmetric but the vacuum is not. This is the condition to obtain spontaneous symmetry breaking. This is exactly the path followed in \cite{Zanzi:2014twa}: spontaneous breakdown of rotational symmetry is useful to justify the diffraction pattern on the screen. For further details about the chameleon-induced collapse of the wave function, the reader is referred to \cite{Zanzi:2014twa, Zanzi:2015kha}.

As far as the organization of this review is concerned, in {Section 2} we will present the standard chameleonic scenario with a discussion of the conformal transformation and of the thin-shell mechanism. In {Section 3} we will further explore the connection between chameleon fields and alternative theories of gravitation. In particular we will analyze {$f(R)$} theories and also the MFM. In the final paragraph we will briefly summarize some concluding remarks.

As already mentioned above, this is a short review about chameleons.
For further
details on the subject the reader is referred to \cite{Mota:2006fz,
Waterhouse:2006wv, Weltman:2008ll, Khoury:2013yy} for reviews and to \cite{Upadhye:2012fz} for a recent summary of some experimental constraints. For a discussion of the stability issue in cosmology the reader is referred to \cite{Roy:2015cna}.

As far as our notation is concerned, we will follow the standard notations of the chameleonic literature and we will call $g_{\mu\nu}$ the E-frame metric, while the J-frame metric will be called $\tilde g_{\mu\nu}$ or, when we consider different sets of matter particles, $g_{\mu\nu}^i$. The only exception will be in the last section where a different notation will be exploited (and explained).

\setcounter{equation}{0}
\section{Standard Chameleonic Theories}
\label{Chameleon}

A chameleon scalar field \cite{Khoury:2003aq, Khoury:2003rn} is a
scalar field coupled to matter (including the baryonic one) with
gravitational strength (or even higher \cite{Mota:2006ed}) and with a mass dependent
on the density of the environment. Before the proposal of
\cite{Khoury:2003aq, Khoury:2003rn}, a discussion of these ideas
had been given in theories with time-varying alpha in
\cite{Mota:2003tc, Mota:2003tm}. Here we are going to consider
only scalar fields, however also chameleonic {{ vector}} fields
have been discussed in the literature \cite{Nelson:2008tn}.

The name ``chameleon'' is due to the peculiar environment-dependent mass of the field.
For example, on very large (\emph{i.e.}, cosmological) length scales,
we know that the matter density is extremely small  and, in this case, the chameleon mechanism is (almost) not operative. Hence the mass of the chameleon
can be of the order of the Hubble constant and the field can roll down the potential on cosmological time scales (for a discussion of ultralight scalar fields in connection to the accelerated expansion of a low redshift Universe see \cite{Copeland:2006wr}).
The careful reader may be worried by this set up, because, for phenomenological reasons, in general it's not possible to introduce in a model a very light scalar field with a generic coupling with matter.
However, this situation is modified if we consider small length scales, for example, \mbox{``this room''}.
In this second case, indeed, the density is higher, the
chameleon mechanism is operative and the field acquires a mass that is large enough to satisfy all experimental bounds on deviations from general
relativity (GR). The chameleon mechanism is a screening mechanism.

To see how this works, let us consider the following scalar-tensor (ST)
theory in the E-frame (see for example \cite{Brax:2004px}):
\begin{equation}
S =\int d^4 x\sqrt{-g} \left\{\frac{M_{Pl}^2R}{2}- \frac{(\partial
\phi)^2}{2} -V(\phi) + \overline{{\cal L}}_m(\psi_m,B^2(\beta
\phi/M_p)g_{\mu\nu})\right\} \label{camstandard}
\end{equation}
where $\phi$ is the chameleon scalar field, $\beta$ is a real
constant that will be discussed later, $M_p$ is the Planck mass
and $V(\phi)$ is the scalar potential. Fermion (matter) fields,
denoted by $\psi_m$, couple conformally to the chameleon through
the $B^2(\beta \phi/M_p )$ dependence of the matter lagrangian
$\overline{{\cal L}}_m$.

A crucial point about the chameleon field is realizing that the
dynamic behavior of the field is not determined only by a
potential but it is governed by an {effective} potential,
$V_{{\rm eff}}$, which depends explicitly on $\rho_m$ (energy
density of non-relativistic matter, {conserved} with respect
to the Einstein frame metric):
\begin{equation}
V_{\rm eff}(\phi)=V(\phi) +\rho_m B(\beta \phi/M_p)\,
\end{equation}

\begin{figure}[H]
\begin{center}
\includegraphics[width=6.5cm]{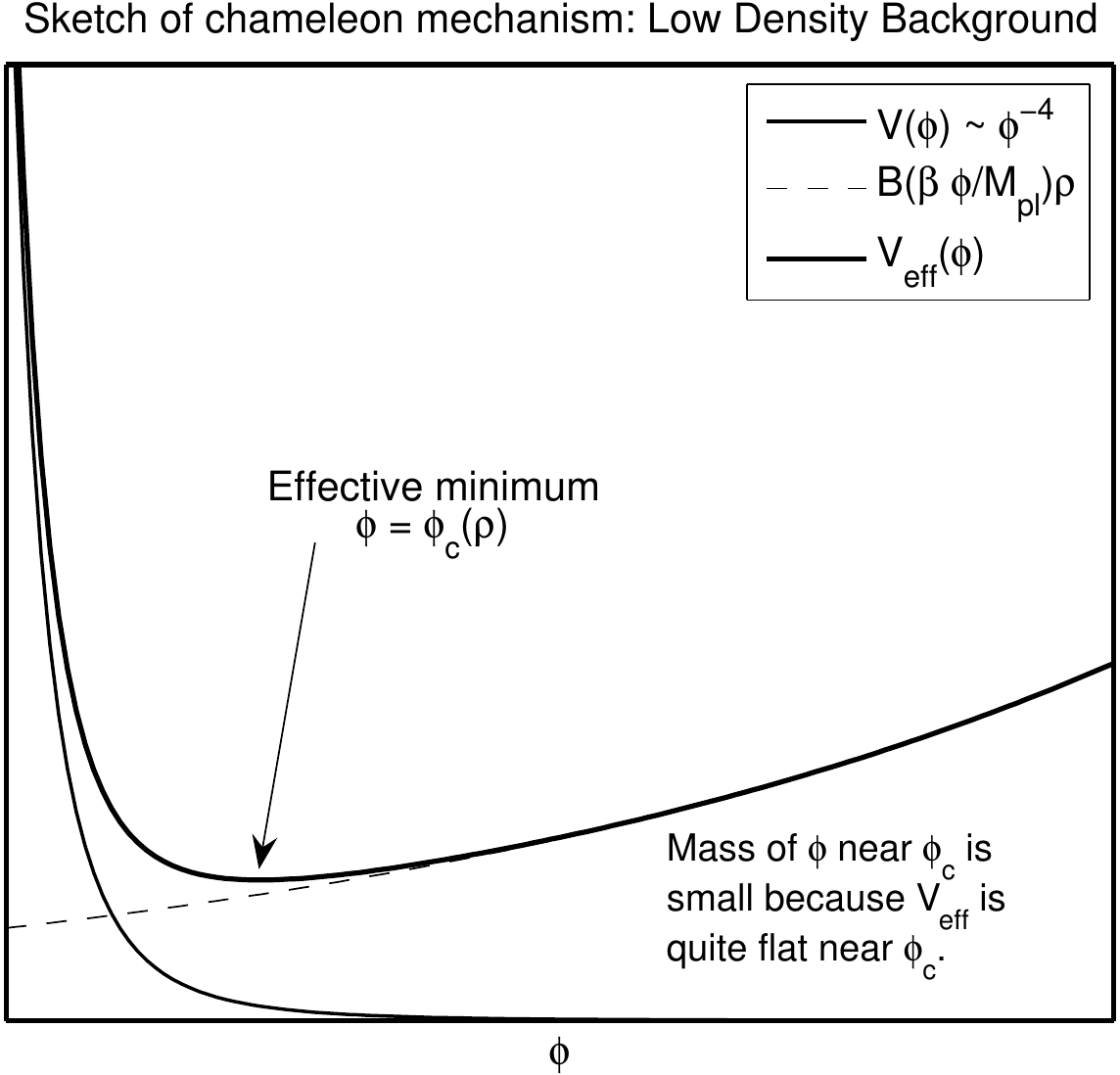}
\includegraphics [width=6.5cm]{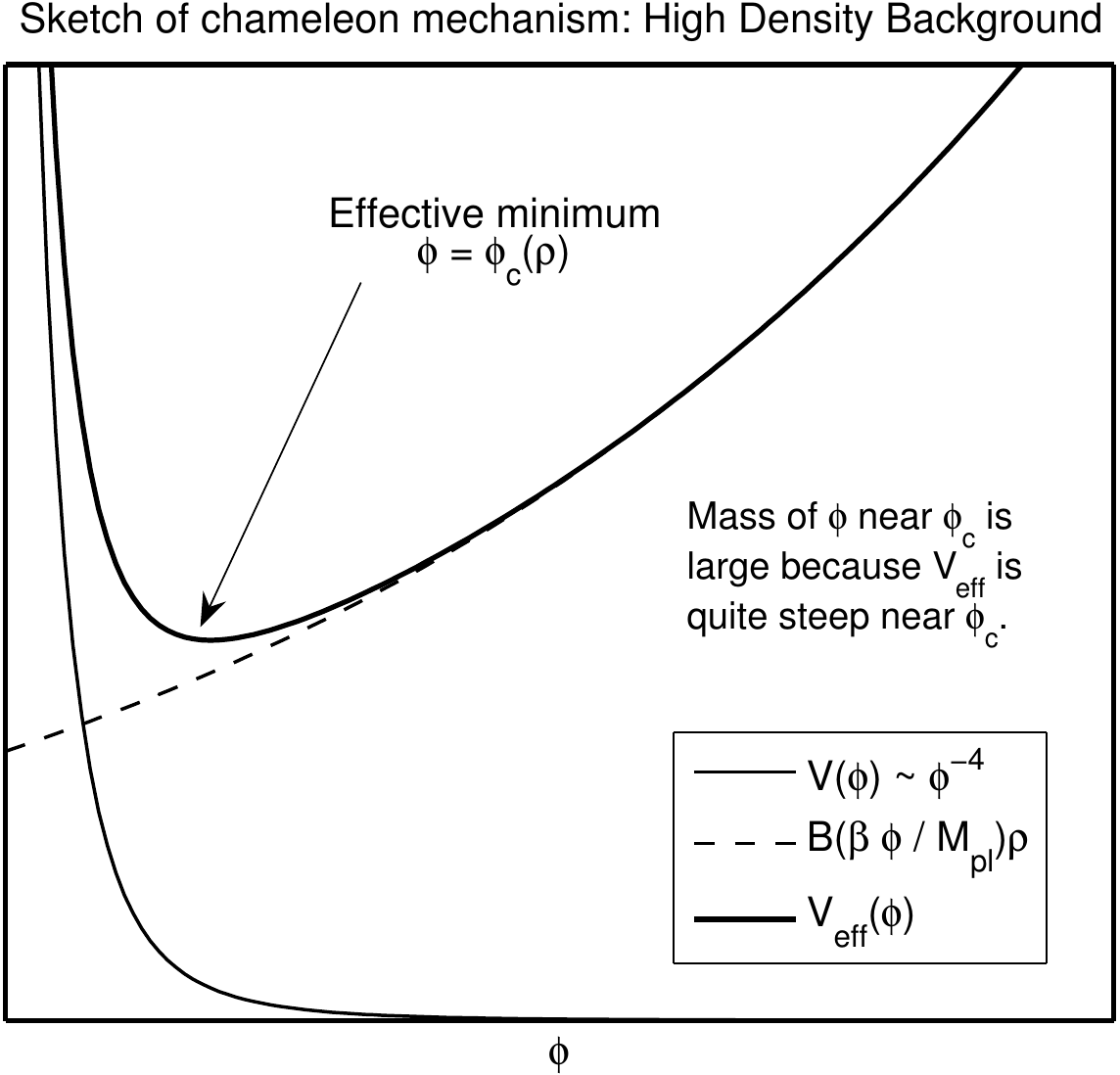}
\end{center}
\caption{{{Chameleon mechanism for a runaway}
potential: $V \sim \phi^{-4}$.  The matter density is small in the left plot and it's large in the right plot (see \cite{Mota:2006fz}).
The mass of the chameleon is an increasing function of the background matter density. These plots can be found in  \cite{Mota:2006fz}. (Reprinted figures with permission from David F. Mota and\mbox{ Douglas J. Shaw}. {\emph{Phys. Rev.} \textbf{2007}, \emph{D75}, 063501.} Copyright (2007) by the { American Physical Society.} \mbox{Source: {{http://journals.}}aps.org/prd/abstract/10.1103/PhysRevD.75.063501})}} \label{FIGphirun}
\end{figure}

A typical example of chameleonic scenario is given by a run-away ``bare'' potential $V(\phi)$ and by a matter branch where
$B(\phi)$ increases with $\phi$. In this way, a ``competition'' between
the two branches of the curve is created and, therefore, the chameleonic environment-dependent mass is obtained. \mbox{The minimum} of the effective potential and the mass of the field in the minimum,
$m^2=V_{,\phi\phi}^{{\rm eff}}$, both depend on $\rho_m$ (see Figure \ref{FIGphirun}) Actually, a run-away bare potential is not strictly necessary if our intention is to construct a chameleon field.

This chameleon mechanism is often related in macroscopic bodies
with the so-called ``thin-shell''.  \mbox{A body} has a {{thin-shell}} if $\phi$ is approximately constant everywhere
inside the body but in a small region (the thin-shell) near the surface of the
body, where large ($\mathcal{O}(1)$) variations of $\phi$ can
occur.  Inside a body
with a thin-shell $\vec{\nabla}\phi$ vanishes everywhere apart
from a thin superficial layer. The force
mediated by $\phi$ is proportional to $\vec{\nabla}\phi$,
consequently, only the thin-shell feels and contributes to the chameleon-mediated ``fifth force''.

Needless to say, experimental bounds on the coupling between the chameleon
and matter must be taken into account. For this purpose, the thin-shell effect plays a crucial role. For example: in the solar system, the standard chameleon field can be
very light and, hence, it can mediate a long-range force whose phenomenological consequences might be unacceptable because the limits on
such forces are very tight.  However, since the chameleon is coupled only
to a small fraction of the matter in large bodies (\emph{i.e.},
that fraction in the thin-shell), the chameleon force between the
Sun and the planets is very weak. We infer that the ``dangerous'' bounds
on long-range forces are faced \cite{Khoury:2003aq,
Khoury:2003rn}. As we will see, the thin-shell mechanism is related to the non-linear nature of the chameleon theories.

\subsection{The Typical Set Up}

In the original proposal \cite{Khoury:2003aq, {Khoury:2003rn}},
the chameleon mechanism is obtained by giving
the scalar field a potential $V(\phi)$ and a coupling to
matter described by a function $B(\beta \phi /M_{pl})\rho$; where $\rho$ is the local
 matter density and $B$ is a function to be specified later. The
potential and the coupling to matter create an {
effective} potential: $V^{eff}(\phi) =
V(\phi) + B(\beta \phi /M_{pl})\rho$.  The values $\phi$ takes at
the minima of this effective potential are environment-dependent.

If $\phi = \phi_c$ is a critical point of the effective potential, \emph{i.e.}, $V^{eff}_{,\phi}(\phi_c)=0$, then the
effective ``mass'' ($m_c$) of the field about $\phi_{c}$
will be obtained evaluating the second derivative of the effective potential ($m_c^2 =
V^{eff}_{,\phi \phi}(\phi_c)$).  If $V(\phi)$ is neither constant,
linear nor quadratic in $\phi$ then $V_{,\phi \phi}(\phi_c)$, and
hence the mass $m_{c}$, will depend on $\phi_c$.  Since $\phi_c$
depends on the matter density, also the effective mass
will be environment-dependent.  A potential $V(\phi)$ with these properties
will lead us to non-linear field equations for $\phi$.

\begin{figure}[H]
\begin{center}
\includegraphics[width=6.5cm]{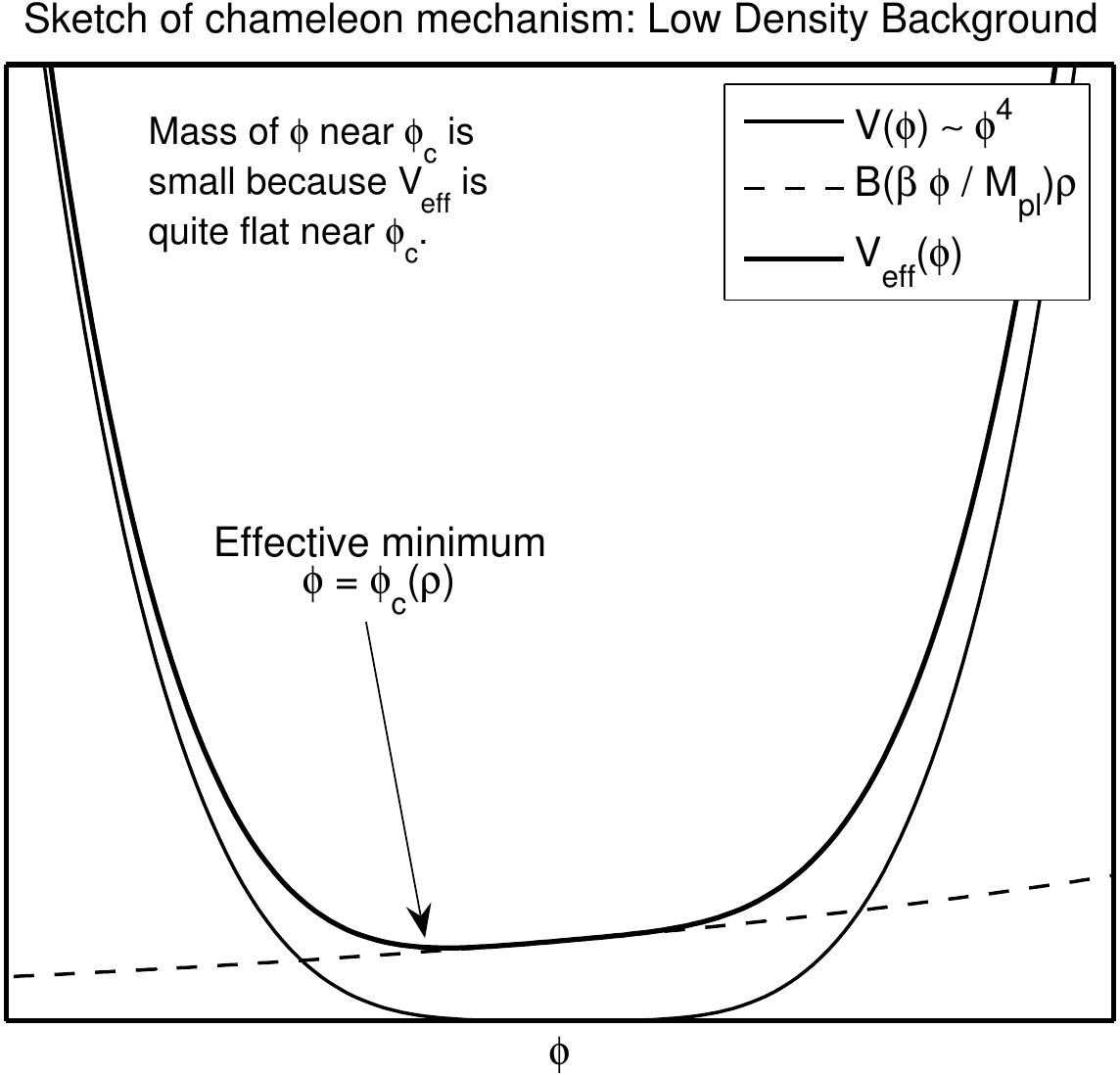}
\includegraphics[width=6.5cm]{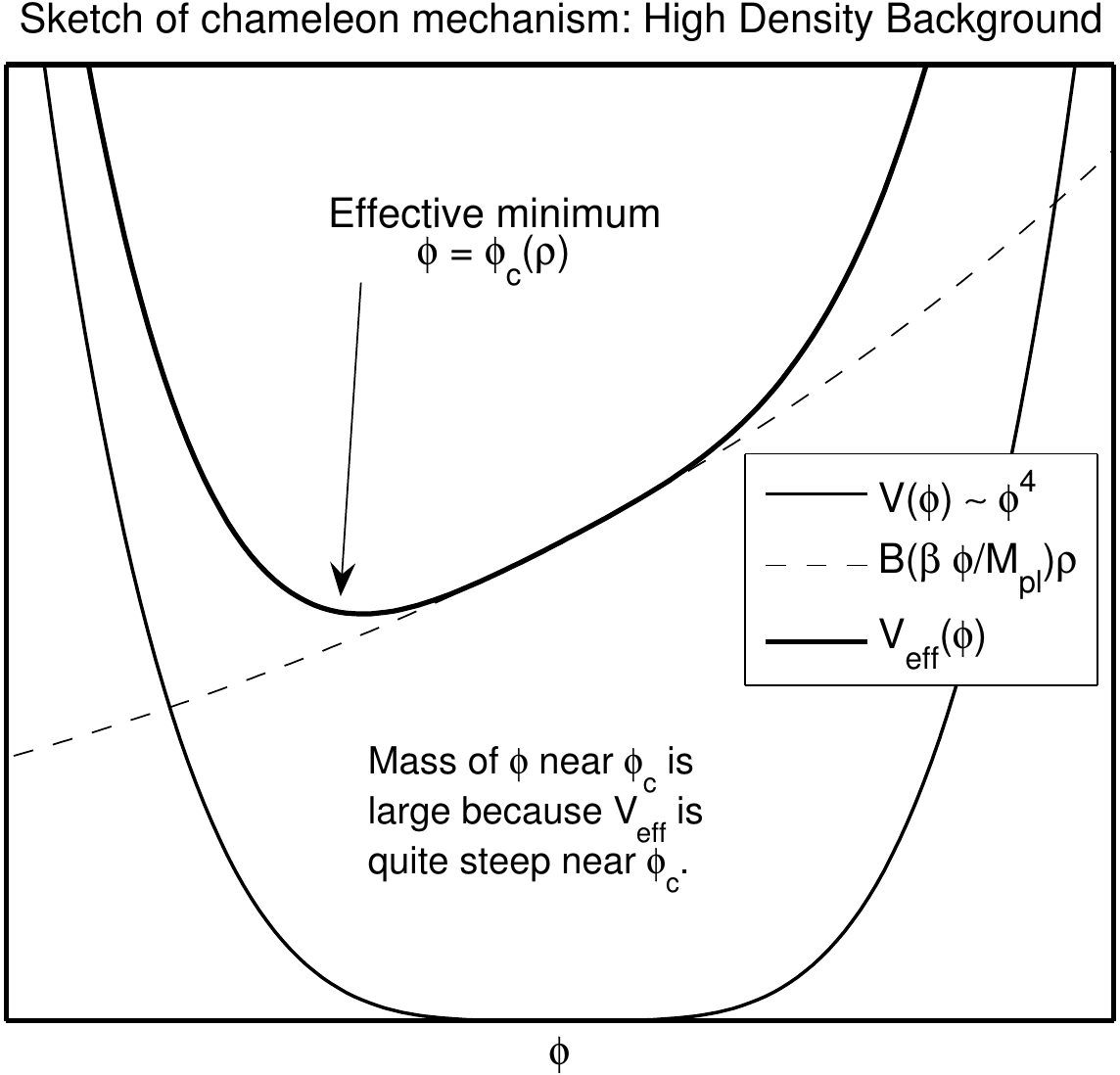}
\end{center}
\caption{{Chameleon mechanism for a potential
with a minimum at $\phi=0$: \mbox{$V \sim \phi^{4}$.}  The left plot corresponds to small matter density and the plot to the right corresponds to large matter density (see \cite{Mota:2006fz}). Once again, the mass of the chameleon is growing with the
matter density. These plots can be found \mbox{in  \cite{Mota:2006fz}.} (Reprinted figures with permission from David F. Mota and Douglas J. Shaw. {\mbox{\emph{Phys. Rev.} \textbf{2007},} {\emph{D75}}, 063501.} Copyright (2007) by the American Physical Society. Source: {{http://journals.aps.org/prd/abstract/10.1103/PhysRevD.75.063501})}} \label{FIGphimin}}
\end{figure}

For a scalar field theory to be chameleonic, the effective
mass of the scalar must be an {{ increasing}} function of the matter density and, therefore,
$V_{,\phi\phi\phi}(\phi_c)/V_{,\phi}(\phi_c) > 0$.  As already mentioned above, it is {{not}} necessary for either
$V(\phi)$ or $B(\beta \phi /M_{pl})$ to have any minima
themselves if our intention is to construct a minimum in the effective potential.
The chameleon mechanism is summarized in \mbox{Figures \ref{FIGphirun} and \ref{FIGphimin}.}

These plots require some additional remarks.
In Figure \ref{FIGphirun} we have a run-away potential, but $V^{eff}$ {{does}}
have a minimum due to the presence of the matter branch and the
value of $\phi$ in that minimum is density dependent.  In Figure
\ref{FIGphimin} the potential is $V \simeq \phi^4$ and
so it {does} have a minimum at $\phi=0$ (it is not run-away), however, the minimum
of the effective potential does {not} coincide
with that of $V$ because the chameleonic coupling to matter creates a shift of the ground state.

Typically when a scalar field $\phi$ is coupled to
matter, one of the effects of the coupling is a non-trivial $\phi$-dependence of the mass $m$ of the matter particle (\emph{i.e.}, the
coupling is encoded into a mass-varying term, no matter whether we
consider this coupling as a classical one or as a result of
quantum corrections). The $\phi$-dependence of $m$ can be written as \be m(\phi) = m_0 B\left(\frac{\beta
\phi}{M_{pl}}\right) \ee where $M_{pl}$ is the Planck mass and
$m_0$ is some constant with units of mass whose definition
will depend on the choice of the function $B\left(\frac{\beta
\phi}{M_{pl}}\right)$. $\beta$ defines the strength of the
coupling of $\phi$ to matter. In general, a consequence of a $\phi$-dependent mass of a matter particle is a $\phi$-dependence of the rest-mass density of this particle: \be \rho(\phi)=\rho_0 B\left(\frac{\beta
\phi}{M_{pl}}\right) \ee

The coupling of $\phi$ to the local
energy density of this particle species is given by $\partial
\rho(\phi) / \partial \phi$ which is: \be \frac{\partial
\rho(\phi)}{\partial \phi} = C'\left(\frac{\beta
\phi}{M_{pl}}\right)\frac{\beta \rho(\phi)}{M_{pl}}  \ee where
$C(x) = \ln B(x)$ and $C'(x) = dC(x)/dx$.

If we linearize $C(\beta \phi/M_{pl})$ we find: \be
C\left(\frac{\beta \phi}{M_{pl}}\right) \approx C(0) + \frac{\beta
C'(0) \phi}{M_{pl}} \ee

This truncation is acceptable granted that $(C''(0)/C'(0)) \beta \phi /M_{pl} \ll 1$.

If $C'(0) \neq 0$, we can use the freedom in the definition
of $\beta$ to set $C'(0)=1$ and, in this case, $\beta$ parametrizes
the strength of the chameleon-to-matter coupling.

For example: a particular choice for $B$ that has been discussed
in the literature (\cite{Khoury:2003aq, Khoury:2003rn,
Brax:2004qh}) is \mbox{$B=e^{k \phi /M_{pl}}$} for some $k$. It follows
that \mbox{$C = k \phi / M_{pl}$,} and so we choose $\beta = k$ which
ensures $C'(0)=1$, $C''(0)=0$.

\subsection{One Possible Lagrangian}

The standard Einstein-Hilbert action is given by
\begin{displaymath}
S_{\text{EH}}=\int d^4 x\sqrt{-g}\frac{1}{16\pi G}R=\int d^4
x\sqrt{-g}\frac{M_\text{pl}^2}{2}R
\end{displaymath}

Let us introduce a scalar field $\phi$ with potential $V(\phi)$ and
action
\begin{displaymath}
S_\phi=-\int d^4
x\sqrt{-g}\left\{\frac{1}{2}(\partial\phi)^2+V(\phi)\right\}
\end{displaymath}

We now define $ -{\cal L}_m\equiv \sqrt{-g} \cdot \overline{{\cal
L}}_m $ and we introduce a set of matter fields
$\psi_\text{m}^{(i)}$ with action
\begin{displaymath}
S_\text{m}=-\int d^4 x
\mathcal{L}_\text{m}\left(\psi_\text{m}^{(i)},g_{\mu\nu}^{(i)}\right)
\end{displaymath}
which are coupled to $\phi$ by the conformal transformation that,
for example, we write as:
\begin{equation}
g_{\mu\nu}^{(i)}\equiv e^{2\beta_i\phi/M_\text{pl}}g_{\mu\nu}
\label{conformalrelation}
\end{equation}
where $\beta_i$ are dimensionless coupling constants, in principle
one for each matter species (\emph{i.e.}, \mbox{multi-metric theory).}

The total action which takes into account gravity, $\phi$ and matter is thus
written as Equation (\ref{camstandard}), namely:
\begin{equation}
\label{total_action}
\begin{aligned}
S&=\int d^4 x\sqrt{-g}\left\{\frac{M_\text{pl}^2}{2}R-\frac{1}{2}\left(\partial\phi\right)^2-V(\phi)\right\}-\int d^4 x \mathcal{L}_\text{m}\left(\psi_\text{m}^{(i)},g_{\mu\nu}^{(i)}\right)\\
&=\int d^4
x\sqrt{-g}\left\{\frac{M_\text{pl}^2}{2}R-\frac{1}{2}\nabla_\mu\phi\nabla^\mu\phi-V(\phi)-\frac{1}{\sqrt{-g}}\mathcal{L}_\text{m}\left(\psi_\text{m}^{(i)},g_{\mu\nu}^{(i)}\right)\right\}
\end{aligned}
\end{equation}

Variation with respect to $\phi$ allows us to obtain the equation
of motion for $\phi$. Following \cite{Waterhouse:2006wv} we write:
\begin{align*}
\delta S&=\int d^4 x\sqrt{-g}\left\{-\nabla_\mu\phi\delta\nabla^\mu\phi-V_{,\phi}(\phi)\delta\phi-\frac{1}{\sqrt{-g}}\frac{\partial \mathcal{L}_\text{m}}{\partial \phi}\delta\phi\right\}\\
&=\int d^4 x\sqrt{-g}\left\{-\nabla_\mu\phi\nabla^\mu\delta\phi-V_{,\phi}(\phi)\delta\phi-\sum_i\frac{1}{\sqrt{-g}}\frac{\partial\mathcal{L}_\text{m}}{\partial g_{\mu\nu}^{(i)}}\frac{\partial g_{\mu\nu}^{(i)}}{\partial\phi}\delta\phi\right\}\\
&=\int d^4 x\sqrt{-g}\left\{\left(\nabla_\mu\nabla^\mu\phi\right)\delta\phi-V_{,\phi}(\phi)\delta\phi-\sum_i\frac{1}{\sqrt{-g}}\frac{\partial\mathcal{L}_\text{m}}{\partial g_{\mu\nu}^{(i)}}\frac{2\beta_i}{M_\text{pl}}g_{\mu\nu}^{(i)}\delta\phi\right\}\\
&=\int d^4
x\sqrt{-g}\left\{\nabla^2\phi-V_{,\phi}(\phi)-\sum_i\frac{1}{\sqrt{-g}}\frac{\partial\mathcal{L}_\text{m}}{\partial
g_{\mu\nu}^{(i)}}\frac{2\beta_i}{M_\text{pl}}g_{\mu\nu}^{(i)}\right\}\delta\phi
,
\end{align*}
where on the first step we exploited the symmetry of the metric to get
$\delta\left(-\frac{1}{2}\nabla_\mu\phi\nabla^\mu\phi\right)=-\nabla_\mu\phi\delta\nabla^\mu\phi$.

On the second step we used the commutativity of differentiation
with variation to obtain \mbox{$\delta\nabla^\mu\phi=\nabla^\mu\delta\phi$.}
On the third step, an integration by parts has been performed by applying the divergence theorem to
$\sqrt{-g}\left(\nabla^\mu\phi\right)\delta\phi$ and assuming that
$\nabla^\mu\phi\rightarrow 0$ or $\delta\phi\rightarrow 0$ at
spacetime infinity or boundaries.

From $\delta S=0$ we obtain the field equation:
\begin{equation}
\nabla^2\phi=V_{,\phi}(\phi)+\sum_i\frac{1}{\sqrt{-g}}\frac{\partial\mathcal{L}_\text{m}}{\partial
g_{\mu\nu}^{(i)}}\frac{2\beta_i}{M_\text{pl}}g_{\mu\nu}^{(i)}
\label{firsteom}
\end{equation}

\subsection{Matter Energy Density}

In this subsection, following \cite{Waterhouse:2006wv}, we will further analyze the matter energy density.
We will fix $i$ and we will write $\tilde X$ instead of $X_{(i)}$.

\label{Defining the energy density of matter}

\subsubsection{Jordan Frame}
If we assume that the matter fields $\psi_\text{m}^{(i)}$ do not
interact with each other, each energy-momentum tensor
\begin{displaymath}
\tilde{T}^{\mu\nu}\equiv-\frac{2}{\sqrt{-\tilde{g}}}\frac{\partial\mathcal{L}_\text{m}}{\partial\tilde{g}_{\mu\nu}}
\end{displaymath}
is conserved in the Jordan frame. Namely,
\begin{equation}
\widetilde{\nabla}_{\nu}\tilde{T}^{\mu\nu}=0
\label{conservationequation}
\end{equation}

Let us assume the matter to be a perfect isentropic fluid with
$\tilde{p}=w_i\tilde{\rho}$.  Then
\begin{equation}
\tilde{T}^{\mu\nu}\tilde{g}_{\mu\nu}=-\tilde{\rho}+3\tilde{p}=-\left(1-3w_i\right)\tilde{\rho}
\label{densityandpressure}
\end{equation}

\subsubsection{Einstein Frame}

Let us impose, without loss of generality, a
Friedmann-Lema\^itre-Robertson-Walker (FLRW) background metric.
The energy density $\rho$ in the Einstein frame (the frame
corresponding to $g_{\mu\nu}$) is conformally
related to $\tilde{\rho}$ and it satisfies the standard continuity
equation $\rho\propto a^{-3\left(1+w_i\right)}$.

Now we collect some useful formulas: \beq \tilde a \equiv e^{\beta_i \phi
/M_{pl}}a \eeq
\beq \tilde{g}_{\mu\nu}=e^{2 \beta_i \phi/M_{pl}}
g_{\mu\nu}=diag(-e^{2\beta_i\phi/M_{pl}},\tilde{a}^2,\tilde{a}^2,\tilde{a}^2);
\eeq \beq \tilde{g}^{\mu\nu}=diag(-e^{-2\beta_i\phi/M_{pl}},\tilde{a}^{-2},\tilde{a}^{-2},\tilde{a}^{-2}).
\eeq

Let us compute the Christoffel symbols $\tilde \Gamma$ (see
\cite{Waterhouse:2006wv}) and let us use them to expand the conservation
Equation (\ref{conservationequation}) with index $\mu=0$. When we differentiate we keep $\phi$ fixed
and we vary the scale factor $a$:
\begin{align*}
0&=\widetilde{\nabla}_{\nu}\tilde{T}^{0\nu}\\
&=\tilde{T}_{,\nu}^{0\nu}+\tilde{\Gamma}_{\sigma\nu}^{0}\tilde{T}^{\sigma\nu}+\tilde{\Gamma}_{\sigma\nu}^{\nu}\tilde{T}^{0\sigma}\\
&=e^{-2\beta_i\phi/M_\text{pl}}\left(\tilde{\rho}_{,0}+3\left(1+w_i\right)\tilde{a}^{-1}\tilde{a}_{,0}\tilde{\rho}\right)
\end{align*}

Thus, multiplying by
$e^{2\beta_i\phi/M_\text{pl}}\tilde{a}^{3\left(1+w_i\right)}$:
\begin{align*}
0&=\tilde{a}^{3\left(1+w_i\right)}\tilde{\rho}_{,0}+3\left(1+w_i\right)\tilde{a}^{3\left(1+w_i\right)-1}\tilde{a}_{,0}\tilde{\rho}\\
&=\left(\tilde{a}^{3\left(1+w_i\right)}\tilde{\rho}\right)_{,0}\\
&=\left(a^{3\left(1+w_i\right)}e^{3\left(1+w_i\right)\beta_i\phi/M_\text{pl}}\tilde{\rho}\right)_{,0}
\end{align*}

In other words, in the Einstein frame, the quantity
\begin{equation}
\label{Einstein_density} \rho\equiv
e^{3\left(1+w_i\right)\beta_i\phi/M_\text{pl}}\tilde{\rho}
\end{equation}
is the energy density of matter, because it satisfies the continuity
equation $\rho\propto a^{-3\left(1+w_i\right)}$.

The Einstein-frame energy
density for each species $i$ is obtained restoring the $i$ subscripts and exploiting {Equation}
(\ref{densityandpressure}):
\begin{equation}
\begin{aligned}
\rho_i&=-e^{3\left(1+w_i\right)\beta_i\phi/M_\text{pl}}\frac{1}{1-3w_i}T_{(i)}^{\mu\nu}g_{\mu\nu}^{(i)}\\
&=e^{3\left(1+w_i\right)\beta_i\phi/M_\text{pl}}\frac{1}{1-3w_i}\frac{2}{\sqrt{-g^{(i)}}}\frac{\partial\mathcal{L}_\text{m}}{\partial g_{\mu\nu}^{(i)}}g_{\mu\nu}^{(i)}\\
&=e^{3\left(1+w_i\right)\beta_i\phi/M_\text{pl}}\frac{1}{1-3w_i}\frac{2}{\sqrt{e^{8\beta_i\phi/M_\text{pl}}\left(-g\right)}}\frac{\partial\mathcal{L}_\text{m}}{\partial g_{\mu\nu}^{(i)}}g_{\mu\nu}^{(i)}\\
&=e^{-\left(1-3w_i\right)\beta_i\phi/M_\text{pl}}\frac{1}{1-3w_i}\frac{2}{\sqrt{-g}}\frac{\partial\mathcal{L}_\text{m}}{\partial
g_{\mu\nu}^{(i)}}g_{\mu\nu}^{(i)}
\end{aligned}
\label{Einstein_energydensity}
\end{equation}

\subsection{The Equation of Motion and the Potential}
\label{sectioncontainingbetaunity}

 If we substitute {Equation}
(\ref{Einstein_energydensity}) into {Equation} (\ref{firsteom}), we
obtain an equation of motion with an \mbox{explicit $\phi$-dependence:}
\begin{equation}
\nabla^2\phi=V_{,\phi}(\phi)+\sum_i\left(1-3w_i\right)\frac{\beta_i}{M_\text{pl}}\rho_i
e^{\left(1-3w_i\right)\beta_i\phi/M_\text{pl}} \label{confronto}
\end{equation}

We infer that the dynamical behaviour of $\phi$ is summarized by an
effective potential
\begin{equation}
\label{veff} V_{\text{eff}}(\phi)\equiv V(\phi)+\sum_i\rho_i
e^{\left(1-3w_i\right)\beta_i\phi/M_\text{pl}}
\end{equation}

Then the chameleon equation of motion in the Einstein frame is
simply
\begin{equation}
\nabla^2\phi=V_{\text{eff},\phi}(\phi) \label{conciseeom}
\end{equation}

We can also write the equations in a slightly more general way:
\be -\square \phi = V_{,\phi}(\phi) + \frac{\beta
B^{\prime}(\beta\phi/M_{pl})(\rho-3P)}{M_{pl}}
\label{confronto2}\ee

The Lagrangian {Equation} (\ref{camstandard}) should {{not}} be viewed as
specifying the {{only}} way in which $\phi$ can be coupled to
matter. However, despite the fact that many different Lagrangians
are possible, it is almost always the case that the field equation
for $\phi$ takes a form very similar to the one given above.

Remarkably, highly non-linear
self-interaction potentials are necessarily present for the chameleon. The non-linear nature of the theory
is a major obstacle if our intention is to obtain analytical solutions of the field equations and this problem is particularly acute
with highly inhomogeneous matter
density.  A linear approximation may
lead us to wrong conclusions about fifth-force experiments.
When the non-linearities are properly taken into account, the chameleon
mechanism becomes much \mbox{stronger \cite{Mota:2006ed,Mota:2006fz, Waterhouse:2006wv}} and this effect gives us the chance to construct models with light cosmological scalars
coupled to matter much \emph{more} strongly than gravity ($\beta
\gg 1$).

To proceed further, we discuss some examples of self-interaction potentials that have been considered in the literature.
Needless to say, $V$ is non-linear and non-quadratic. One example is given by the Ratra-Peebles
potential, $V(\phi)= M^4(M/\phi)^{n}$, where $M$ is some mass
scale and $n > 0$; chameleon fields have also been studied in the
context of $V(\phi)= k\phi^4 / 4!$ (see \cite{Gubser:2004uf}). Let us consider this potential: \be V(\phi) = \lambda M^4 (M/\phi)^{n} \ee where $n$ can
be positive or negative and $\lambda > 0$.  If $n \neq -4$ then we
can scale $M$ so that $\lambda = 1$.
\mbox{When $n=-4$,} $M$ drops out and we have a $\phi^4$ theory. When
$n>0$ this is just the Ratra-Peebles potential. When $n\neq-4$,
our choice of potential comes from
an expansion, for small $(\hat{M}/\phi)^n$, of another potential
$W(\phi) = \hat{M}^4 f((\hat{M}/\phi)^n)$ where $f$ is some
function. We could then write: \be W \sim \hat{M}^4 f(0) +
\hat{M}^4 f'(0) \left(\frac{\hat{M}}{\phi}\right)^{n} \ee where
$\hat{M}$ is some mass-scale. We define $M$ so that the second
term on the right hand side of the above expression reads $M^4
(M/\phi)^n$. The first term on the right hand then plays the role
of a CC \mbox{$\hat{M}^4f(0) \approx \rho_{\Lambda}$.}
If $f(0)$ and $f'(0)$ are $\mathcal{O}(1)$, then the typical ansatz in the chameleonic literature is \mbox{$M \approx \hat{M} \approx
\left(\rho_{\Lambda}\right)^{1/4} \approx (0.1\,
{\mathrm{mm}})^{-1}$. }In other words, the CC scale is inserted by-hand (as already mentioned above this ansatz is not present in \cite{Zanzi:2010rs} where no fine-tuning has been introduced in the model).

Consequently, we are led to the following field equations:
\begin{eqnarray}
- \square \phi = -n \lambda M^3 \left(\frac{M}{\phi}\right)^{n+1}
  +\frac{\beta (\rho+\omega P)}{M_{pl}}
\label{micro}
\end{eqnarray}

In order to render the chameleon mechanism operative, the potential gradient
term \mbox{$V_{,\phi} = -n\lambda M^3 (M/\phi)^{n+1}$} and the matter
coupling term $\beta (\rho + \omega P) / M_{pl}$ must have
opposite signs. It is usually the case that $\beta > 0$ and
$P/\rho \ll 1$. If $n > 0$ we must therefore have $\phi > 0$.  In
theories with $n < 0$ we must have $\phi < 0$ and $n=-2p$ where
$p$ is a positive integer. Another condition that must be satisfied is that the mass squared of the chameleon $ m_{c}^2 = V_{,\phi \phi}$
must be positive and $\phi$-dependent. \mbox{These conditions} mean
that we must exclude the region $-2 \geq n \geq 0$. Interestingly, for $n=-2$,
\mbox{$n=-1$ or} $n=0$ the field equations would be
linear.

\subsection{The Thin-Shell Effect}

In this {(sub)section}, following \cite{Mota:2006fz}, we will
analyze the chameleonic field in the presence of a single body and we
will discuss the solutions to Equation (\ref{micro}) in three
different regimes that will be quantitatively defined later,
namely:
\begin{itemize}
\item A {{ linear}} regime, \emph{i.e.}, a linear approximation to Equation (\ref{micro});  \vspace{-9pt} \item {{ Pseudo-linear}} regime: in this case it is
not possible to linearize Equation (\ref{micro}), but we can construct
multiple linearizations valid in different regions;  \vspace{-9pt} \item {{
Non-linear} }regime: when linear and pseudo-linear approximations
are not valid, the chameleon will have a true non-linear behaviour near the
surface of the object. These non-linearities are related to the
presence of the thin-shell mechanism.
\end{itemize}

Let us consider a spherically symmetric object with
uniform density $\rho_c$ and radius $R$ in a background
with uniform density $\rho_b$. We will work with a Minkowskian (at leading order) spacetime. Under these assumptions $\square
\rightarrow -\nabla^2$ where $\nabla^2 \phi = r^{-2} (r^2
\phi^{\prime})^{\prime}$; $\phi^{\prime} = {\mathrm{d}}\phi /
{\mathrm{d}}r$. Inside the body ($r <
R$) we assume spherical symmetry and, therefore, $\phi$ obeys:
\begin{eqnarray}
\frac{d^2 \phi}{dr^2} + \frac{2}{r}\frac{d \phi}{dr} = -n\lambda
M^3 \left(\frac{M}{\phi}\right)^{n+1} +\frac{\beta
\rho_c}{M_{pl}} \label{inbody}
\end{eqnarray}

Outside the body $(r>R)$ we have:
\begin{eqnarray}
\frac{d^2 \phi}{dr^2} + \frac{2}{r}\frac{d \phi}{dr} = -n\lambda
M^3 \left(\frac{M}{\phi}\right)^{n+1} + \frac{\beta
\rho_b}{M_{pl}} \label{outbody}
\end{eqnarray}

The right hand side of Equation (\ref{inbody}) vanishes in the minimum $V_{eff}$ inside the body, namely when $\phi =
\phi_c$ where $$ \phi_{c} = M\left(\frac{\beta \rho_c}{n\lambda
M_{pl}M^3}\right)^{-\frac{1}{n+1}} $$

Similarly, the right hand side of Equation (\ref{outbody})
vanishes in the minimum of $V_{eff}$ outside the body, namely when $\phi=\phi_b$ where $$ \phi_{b} = M\left(\frac{\beta
\rho_b}{n\lambda M_{pl}M^3}\right)^{-\frac{1}{n+1}} $$

For
large $r$ we must have $\phi \approx \phi_b$. The effective mass
is given by:
\begin{eqnarray}
m^2_{\phi}(\phi) = V^{eff}_{,\phi\phi}(\phi) = n(n+1)\lambda M^2
\left(\frac{M}{\phi}\right)^{n+2} \label{effmass}
\end{eqnarray}

We define $m_c = m_{\phi}(\phi_c)$ and $m_b = m_{\phi}(\phi_b)$.
We shall see below that the larger the quantity $m_c R$ gets, the
more likely it is that a body will have a thin-shell. Throughout
this section we will require these boundary conditions:
\begin{eqnarray*}
\left.\frac{d\phi}{d r}\right\vert_{r=0} = 0\quad {\mathrm{and}}
\quad \left.\frac{d\phi}{d r}\right\vert_{r=\infty} = 0
\end{eqnarray*}

\subsubsection{Linear Regime}

We assume that it is a valid
approximation to linearize the equations of motion for $\phi$
about the value of $\phi$ in the far background, $\phi_{b}$.
The linear approximation is valid granted that
certain conditions are satisfied. We will summarize these
conditions later in this section. We
write $\phi = \phi_b + \phi_1$ and the linearized field equations
are:
\begin{equation}
\frac{d^2 \phi_1}{d r^2} + \frac{2}{r}\frac{d \phi_1}{d r} = -n
M^3 \left(\frac{M}{\phi_b}\right)^{n+1} + m^2_{b}\phi_1 +
\frac{\beta (\rho_{c}-\rho_{b})}{M_{pl}}H(R-r) + \frac{\beta
\rho_b}{M_{pl}}
\end{equation}
where $H(R-r)$ is the Heaviside function: $H(x)=1$, $x\geq 0$, and
$H(x)=0$, $x<0$. This linearization of the potential is
acceptable granted that $$ \frac{V_{,\phi
\phi}(\phi_b)\phi_1}{V_{,\phi}(\phi_b)} < 1$$ and this gives
$\vert \phi_1/\phi_b \vert < \vert n+1 \vert^{-1}$. Moreover, for
this linearization to remain valid as $r \rightarrow \infty$, we
need $\phi_1 \rightarrow 0$, which implies that $$ n M^3
\left(\frac{M}{\phi_b}\right)^{n+1} = \frac{\beta \rho_b}{M_{pl}}
$$

The solution to the field equations outside the body ($r>R$) is
 $$
\phi_1 = \frac{\beta \Delta \rho_c}{M_{pl}m_b^2}\frac{e^{m_b(R-
r)}}{m_{b}r}\left(\frac{\tanh(m_b R)-m_bR}{1+\tanh(m_bR)}\right)
$$ where we defined $\Delta \rho_c = \rho_c-\rho_b$, while inside the body ($r<R$) $\phi_1$ is given by $$ \phi_{1} =
-\frac{\beta \Delta \rho_{c}}{m_{b}^2 M_{pl}} + \frac{\beta \Delta
\rho_c}{M_{pl}m_b^2}\frac{\left(1+m_b R\right)e^{-m_bR}\sinh(m_b
r)}{m_b r} $$

The largest value of $\vert\phi_1/\phi_b\vert$
occurs at $r=0$ and hence
$\vert \phi_{1}(r=0)/\phi_{b} \vert < \vert n+1 \vert^{-1}$ is required for an acceptable linear approximation. This
requirement is equivalent to
$$ \left\vert(1+m_b R)e^{-m_b R}-1\right\vert \frac{\Delta
\rho_{c}}{\rho_b} \sim \frac{1}{2}m_b^2R^2 \frac{\Delta
\rho_{c}}{\rho_b} = \frac{\Delta
\rho_c}{\rho_c}\left(\frac{\rho_b}{\rho_c}\right)^{\frac{1}{n+1}}
\frac{(m_cR)^2}{2}<1 $$
where ``$\sim$'' means ``asymptotically in
the limit $m_b R \rightarrow 0$''. As we can see from this last
formula, if we consider the typical situation where $\rho_b \ll
\rho_c$, then for theories with $n>0$, the lower the density of
the background gets, the better the linear approximation will be
(because $\phi_b$ will be larger). On the contrary, when $n \leq -4$ the
opposite is true.

\subsubsection{Pseudo-Linear Regime \label{singpseu}}

We now assume the existence of (at least) one self-consistent
linearization of the field equations about every point (we
will define these existence conditions later). We do not require
the validity of this linearization {{everywhere}}. Instead we
construct two linearizations of the field equations: the
{{inner} }and the {{outer approximations}} to $\phi$.

The {{inner approximation}} is an asymptotic
approximation to the chameleon that is valid, on the one hand, inside an
isolated body and, on the other hand, near the surface of that body. In general, the
inner approximation will not be valid anymore far from the
body ($r \gg R$).

The second linearization, the {outer approximation}, is an asymptotic approximation to
$\phi$ that is valid for large values of $r$, but, in general, not for $r \sim \mathcal{O}(R)$. We require that it
remains valid as $r \rightarrow \infty$.

The boundary conditions already mentioned above are:
$$ \left.\frac{d \phi}{dr}\right\vert_{r=0} = 0, \qquad \left.\frac{d
\phi}{dr}\right\vert_{r=\infty} = 0
$$

We cannot apply the $r=0$ boundary condition
to the outer-approximation. Similarly the $r=\infty$ boundary
condition will be applicable to the outer-approximation but not to
the inner one.

It is not possible to apply all the boundary conditions to both
approximations and, consequently, there will be undefined constants of integration.
However, these constants can be evaluated if we find an {intermediate} range of values
of $r$ ($r_{out}< r < r_{in}$ say) where the inner
{and} the outer approximations are both valid.

As pointed out in \cite{Mota:2006fz} and references therein,
asymptotic expansions are locally unique. Therefore, if both
the outer and inner approximations are simultaneously valid in
some intermediate region, then they must be equal to each other in
that region.

\vspace{12pt}
\noindent{Inner Approximation}
\vspace{12pt}

Inside the body, $0\leq r \leq R$, the chameleon satisfies:
\begin{eqnarray}
\frac{d^2 \phi}{dr^2} + \frac{2}{r}\frac{d \phi}{dr} = -n\lambda
M^3\left(\frac{M^3}{\phi}\right)^{n+1} + \frac{\beta
\rho_c}{M_{pl}} \label{phiAppev}
\end{eqnarray}

The inner approximation is defined by the assumption $$ n\lambda
M^3\left(\frac{M^3}{\phi}\right)^{n+1} \ll \frac{\beta
\rho_c}{M_{pl}}$$

If we define $$ \phi_c = M\left(\frac{\beta
\rho_c}{n \lambda M_{pl}M^3}\right)^{-\frac{1}{n+1}} $$

We see that the above assumption is equivalent to: $$ \delta(r) :=
\left(\frac{\phi_c}{\phi(r)}\right)^{n+1} \ll 1 $$

We define the
inner approximation by solving Equation (\ref{phiAppev})
 for $\phi$ as an asymptotic expansion in the small parameter
$\delta(0)$. It is noteworthy that $\delta(r)< \delta(0):=\delta$.

Whenever the inner approximation is valid we have:
$$ \phi \sim \phi_0 + \frac{\beta \rho_c r^2}{6 M_{pl}} + \mathcal{O}(\delta):= \bar{\phi}(r)+ \mathcal{O}(\delta)
$$ for $r<R$ with the order $\delta$ term given by
$$ \phi_{\delta}(r) = \frac{\delta\beta \rho_c}{r M_{pl}}\int_{0}^{r}
dr^{\prime} \int_{0}^{r^{\prime}}dr^{\prime \prime}\, r^{\prime
\prime} \left(\frac{\phi_0}{\bar{\phi}(r^{\prime
\prime})}\right)^{n+1}
$$
 $\phi_0$ is an undefined constant of integration to be
determined by matching the inner approximation to the outer one.
In order to guarantee the validity of the inner approximation inside the body we
need:
$$ \frac{\phi_{\delta}(r)}{\bar{\phi}(r)} \ll 1
$$

Outside the body, $r > R$, $\phi$ satisfies:
$$ \frac{d^2 \phi}{dr^2} + \frac{2}{r} \frac{d \phi}{dr} = -n \lambda
M^3 \left(\frac{M^3}{\phi}\right)^{n+1}
$$
where the $\rho$-matter contribution can be safely neglected
because we are considering the solution close to the body.
Whenever $\delta(r) < 1$, the above equation can be solved in the
inner approximation (for an explicit expression of
${\bar{\phi}(r)}$ for $r>R$ see \cite{Mota:2006fz}):
$$ \phi \sim \bar{\phi}(r) + \phi_{\delta}(r)
$$
for $r>R$.

Hence, the inner approximation will be valid simultaneously inside and
outside the body granted that $$
\frac{\phi_{\delta}(r)}{\bar{\phi}(r)} \ll 1 $$ In general, this
requirement will hold only for $r$ smaller than some finite value
of $r$ ($r=r_{in}$).

\vspace{12pt}
\noindent{Outer Approximation}
\vspace{12pt}

When $r$ is very large, the presence of the body
should induce only a small perturbation on $\phi$.
\mbox{If we} assume that $\phi \rightarrow \phi_b$ for $r \rightarrow
\infty$, then the outer approximation is defined by the assumption
\mbox{$\vert (\phi - \phi_b) / \phi_b \vert < 1/\vert n+1 \vert.$ }We infer that: $$ -n \lambda
M^3\left(\frac{M^3}{\phi}\right)^{n+1} \sim -n \lambda
M^3\left(\frac{M^3}{\phi_b}\right)^{n+1} + m_b^2 (\phi-\phi_b) +
O\left((\phi/\phi_b -1)^2\right) $$ where $$ m_b^2 = \lambda
n(n+1) M^2 \left(\frac{M}{\phi}\right)^{n+2} $$ is the mass of the
chameleon in the background.  The assumption $\vert (\phi -
\phi_b) / \phi_b \vert < 1/\vert n+1 \vert$ is basically the
same assumption considered in the linear approximation, with the
difference that, in the pseudo-linear case, it is not required to
hold up to $r=0$ but only for $r
> r_{out}$, where $r_{out}$ is any value of $r$ smaller than
$r_{in}$. In other words, some intermediate region where
both (the inner and the outer) approximations are simultaneously
valid is necessary.

Outside of the body $\phi$ satisfies: $$ \frac{d^2 \phi}{dr^2} +
\frac{2}{r} \frac{d \phi}{dr} = -n \lambda M^3
\left(\frac{M^3}{\phi}\right)^{n+1} + \frac{\beta\rho_b}{M_{pl}}
$$ where the $\rho$-matter contribution cannot be neglected anymore.
In order to guarantee a valid outer approximation when $\{r \rightarrow
\infty,\, \phi \rightarrow \phi_b\}$, it is necessary that $$ n \lambda M^3
\left(\frac{M^3}{\phi_b}\right)^{n+1} =
\frac{\beta\rho_b}{M_{pl}} $$

Solving for $\phi$ in the outer
approximation, we find $\phi \sim \phi^{\ast}$ where:
\begin{eqnarray}
\phi^{\ast} = \phi_b - \frac{\chi e^{-m_b r}}{r} \label{phiast}
\end{eqnarray}
$\chi$ is an unknown constant of integration that will
be determined through the matching procedure.

\vspace{12pt}
\noindent{Matching Procedure}
\vspace{12pt}

Let us suppose that an intermediate region
$r_{out} < r < r_{in}$ exists where the inner and outer
approximations are both valid (we will discuss the existence conditions of
this open set below). \mbox{In other }words, an open set
about some point $r=d$ where both approximations are valid is required. We
must also impose the validity of the inner approximation for all
$r < d$ and of the outer approximation for all $r > d$. In the
intermediate region the uniqueness of the asymptotic expansion implies
$$ \phi \sim \bar{\phi} \sim \phi^{\ast} $$
 $\phi_0$ and $A$
can be evaluated from this last formula and the result is:
\begin{eqnarray}
\phi_0 &=& \phi_b - \frac{\beta \rho_c}{2M_{pl}} \\ \chi &=&
\frac{\beta \rho_c}{3M_{pl}} \label{costanti}
\end{eqnarray}

Now that $\chi$ and $\phi_0$ have been determined, we will
evaluate the existence conditions of an \mbox{intermediate region.}

\newpage
\vspace{12pt}
\noindent{Conditions for Matching}
\vspace{12pt}

In order to have an outer approximation valid for all $r>d$ we need:
$$ \left\vert\frac{(n+1)(\phi^{\ast} - \phi_b)}{\phi_b}\right \vert
\ll 1$$

Using Equations (\ref{costanti}) and (\ref{phiast}) we can see
that this is equivalent to:
\begin{eqnarray}
(m_c R)^2 \frac{R}{3d} \ll
\left(\frac{\rho_c}{\rho_b}\right)^{1/(n+1)} \label{mccond}
\end{eqnarray}

We define $d_{min}$ to be the value of $d$ such that the left hand
side and right hand side of Equation (\ref{mccond}) are equal. An
outer approximation valid at $d$ requires $d > d_{min}$.

In order to have a valid inner approximation, we require
$(\bar{\phi}(r=0)/\phi_c)^{-(n+1)} \ll 1$ or equivalently
\begin{eqnarray} (m_c R)^2 \ll 2\vert n+1 \vert\left((\rho_c/\rho_b)^{1/(n+1)}
-1\right) \label{inhold}
\end{eqnarray}

In order to have the inner approximation ($\phi \sim \bar{\phi}$) valid in
the intermediate region, we require $$ \frac{R^3}{3}\gg
\int^{r}_{0}d r' \int^{r'}_{0} dr^{\prime \prime} r^{\prime
\prime} \left(\frac{\phi_c}{\bar{\phi}(r^{\prime
\prime})}\right)^{n+1} $$ for all $R< r < d$ and also the validity
of Euqation (\ref{inhold}). Granted that this is the case then, for all $r$
in $(R,d)$, we need
\begin{eqnarray}
\frac{R^3}{3}\gg \int^{d}_{R}d r' \int^{d}_{r'} dr^{\prime \prime}
r^{\prime \prime} \left(\frac{\phi_c}{\bar{\phi}(r^{\prime
\prime})}\right)^{n+1}. \label{intexr}
\end{eqnarray}

If both Equations (\ref{inhold}) and (\ref{intexr}) hold, then the
inner approximation will be valid for all $r < d$ (see
\cite{Mota:2006fz} for a more complete discussion).

From Equation (\ref{mccond}) and (\ref{inhold}) we infer that, if our
intention is to guarantee the existence of an intermediate region,
the product $m_c R$ must be small enough. This is a very general
result: whatever will be the potential we choose ($n<-4$, $n>0$ or
$n=-4$), the pseudo-linear approximation will not be valid
everywhere if $m_c R$ is not small enough. In this case,
non-linear effects begin to become important near the centre of
the body.  If we increase $m_c R$, the region where
non-linear effects are relevant moves out from the centre of the
body. If $m_c R$ is large enough, the non-linear nature
of the chameleon potential, $V(\phi)$, is relevant only in a thin
region near the surface of the body, namely the thin-shell (see
\cite{Mota:2006fz}). \mbox{In this} regime the body is very large with
respect to $m_c^{-1}$ and we expect $\phi\simeq\phi_c$ inside the
body. We will discuss this issue in more detail in {Section}
\ref{armonica}, while in {Section} \ref{nonclose}, following
\cite{Mota:2006fz}, we will show that the condition $m_c R
>>1$ implies $\Delta R/R<<1$, where $\Delta R$ is the thickness of
the shell.

\subsubsection{Non-Linear Regime Close to the Body \label{nonclose}}

If our intention is to analyze the evolution of $\phi$ in the thin-shell,
the curvature of the surface of the body can be neglected,
because the thickness of the shell $\Delta R$ is much smaller than
$R$. \mbox{Consequently, we} treat the surface of the body as a flat surface,
with outward normal in the direction of the positive $x$-axis. The
surface of the body is by definition at $x=0$ (\emph{i.e.}, $x=r-R$).
Naturally, we are
interested in physics over length-scales much
smaller than the size of the body, because the shell is much thinner than the length scale of the body.
Hence, we can exploit the
approximation that the body extends to infinity along the $y$ and
$z$ axes and also along the negative $x$ axis. With these
assumptions,  $\phi$ satisfies

$$ \frac{d^2 \phi}{d x^2} = -n\lambda M^3 (M/\phi)^{n+1} + \beta
\rho_c / M_{pl}$$

Our boundary conditions (BCs) are $\phi \rightarrow \phi_c$
and $d \phi / d x \rightarrow 0$ as $x \rightarrow -\infty$.  With
these BCs, the first integral of the above equation is:
\begin{equation}
\frac{1}{2}\left(\frac{d \phi}{dx}\right)^2 \approx \lambda M^4
\left[(M/\phi)^n - (M/\phi_c)^n\right] + \frac{\beta
\rho_c}{M_{pl}} (\phi - \phi_c) \label{phiin}
\end{equation}

Outside of the body, we assume (1) $\phi \rightarrow \phi_b$ as
$x \rightarrow \infty$; and (2) a background with density
$\rho_{b}\ll \rho_{c}$. If $m_c R$ is large enough, we can safely
assume
$$ \left\vert \frac{d^2\phi}{d x^2} \right\vert \gg
\left\vert\frac{2}{r}\frac{d \phi}{dx}\right\vert
$$ then we can ignore the curvature of the surface of the body and, in
$x>0$, we have:
\begin{eqnarray}
\frac{1}{2}\left(\frac{d \phi}{dx}\right)^2 = \lambda
M^4(M/\phi)^{n} - \lambda M^4(M/\phi_{b})^{n} + \frac{\beta
\rho_b}{M_{pl}}(\phi - \phi_{b}) \label{phiout}
\end{eqnarray}

Our assumption that $\vert d^2 \phi / d x^2 \vert \gg (2/r)d \phi
/ d x$ then requires that: $$
\frac{2\sqrt{2}}{m_{\phi}(\phi)r}\sqrt{\frac{n+1}{n}}\left(1-(n+1)\left(\frac{\phi}{\phi_b}\right)^{n}
+ n\left(\frac{\phi}{\phi_{b}}\right)^{n+1}\right)^{1/2} \ll 1 $$

Near the surface of the
body, we expect   $\phi \sim \mathcal{O}(\phi_c)$ granted that the pseudo-linear approximation breaks down and that the
body has a thin-shell. Hence, whenever
$\rho_c \gg \rho_b$, we have $(\phi/\phi_b)^{n} \ll 1$ and
$(\phi/\phi_b)^{n+1} \ll 1$.  The above condition will therefore
be satisfied provided that $m_cR \gg \sqrt{8(n+1)/n}$; this is
generally a weaker condition than the thin-shell conditions. On the surface, at $x=0$,
$\phi$ and ${\mathrm{d}}\phi/{\mathrm{d}}x$ must be both continuous.
If we impose the continuity condition of ${\mathrm{d}}\phi
/{\mathrm{d}}x$ at the surface, we infer \cite{Mota:2006fz}:
$$ \frac{\phi(0)-\phi_c}{\phi_c} = \frac{1}{n}
$$

As already mentioned above, $m_c R>>1$ implies the existence of
a thin-shell (\emph{i.e.}, $\Delta R \ll R$). Let us prove this statement.
Near the
surface of the body, almost all variation in $\phi$ are expected to take place in
a shell of thickness $\Delta R$. We define
$m_{surf}$ by: $$ \int^{0}_{-\infty} dx \frac{d\phi}{dx} =
\phi(x=0)-\phi_c = \phi_c/n = m_{surf}^{-1} \frac{d\phi}{d
x}(x=0) $$
$m_{surf}^{-1}$ is then, approximately, the length
scale over which any variation in $\phi$ dies off. It follows that
$\Delta R \approx m_{surf}^{-1}$. In order to (A) render this shell thin and
(B) safely ignore the curvature of the surface of
the body, we need $\Delta R / R \ll 1$ or equivalently $m_{surf}R
\gg 1$. If we assume $\rho_b \ll \rho_c$ we can write
\cite{Mota:2006fz}: $$ m_{surf}R \approx
\left(\frac{n}{n+1}\right)^{n/2+1}m_c R \sim \mathcal{O}(m_c R)
$$ and so $m_{surf}R \gg 1$ follows from $m_{c}R \gg 1$, and
$\Delta R \sim \mathcal{O}(m_c^{-1})$. $m_{surf}R \gg 1$ will be
automatically satisfied whenever the thin-shell conditions hold.

Whenever $\rho_{c} \gg \rho_{b}$, Equation (\ref{phiout}) will be well-approximated by $$ \frac{1}{2}\left(\frac{d
\phi}{d x}\right)^2 \approx \lambda M^4 (M/\phi)^n $$  near
$r=R$. If we solve this formula with the boundary conditions mentioned above, namely
\mbox{$\phi(x=0) = (1+1/n)\phi_c$} and $\phi/\phi_c \rightarrow
\phi_b/\phi_c$ as $x \rightarrow \infty$, we have
\begin{eqnarray}
 \frac{1}{m_{\phi}(\phi)} \sim \frac{\vert n+2 \vert (r-R)}{\sqrt{2n(n+1)}}  +\left(\frac{n+1}{n}\right)^{n/2+1}\frac{1}{m_{c}} \label{phiclosef}
\end{eqnarray}

Hence, if $r-R$ is large enough, then $m_{\phi}$ (and
therefore also $\phi$) will be independent of $m_{c}$ and consequently also
of $\phi_c$ and $\beta$ at leading order. We further elaborate on
this point in the \mbox{{(sub)section} below.}

Summarizing, (1) the presence of a thin-shell is related to
non-linear effects that are non-negligible near the surface of the
body; (2) a thin-shell exists whenever $m_c R$ is
large enough.

\subsubsection{Non-Linear Regime Far from the Body \label{farphithin}}

Interestingly, even if $m_c R$ is
large enough,
non-linear effects should not be important far from the surface of the
body.  Indeed, for large $r$, $\phi$ should have a functional form
similar to that found in the pseudo-linear approximation.
We will consider here only run-away potentials ($n >0$) and we refer the
reader to \cite{Mota:2006fz} for a more complete discussion.

Away from the surface of the body we expect that
non-linear effects will be negligible and as $r \rightarrow
\infty$ we will have:
$$
\phi \sim \phi^{(0)} = \phi_b -  \frac{\tilde{\chi} e^{-m_b
r}}{r}
$$
for some constant $\tilde{\chi}$ where $\phi_b$ and $m_b$ are the
values of the chameleon and its mass in the background. In $n>0$
theories the potential is singular and we have $\phi^{(0)}>0$
outside the body. The minimum value of $\phi^{(0)}$ outside the
body occurs at $r=R$ and hence
$$
 \tilde{\chi}< \phi_{b}e^{m_b R}R
$$

Typically, $m_b R \ll 1$ and therefore
$$
\tilde{\chi} < \phi_b R
$$

This upper bound on $\tilde{\chi}$ defines a critical form for the
field outside the body:
$$
\phi_{crit} = \phi_b\left(1 - \frac{e^{m_b (R-r)}R}{r}\right)
$$

No matter what occurs inside the body, we must have $\phi
> \phi_{crit}$ outside the body as $r \rightarrow \infty$.
\mbox{Ignoring non-linear} effects, $\phi > \phi_{crit}$ is satisfied by
all bodies that satisfy the conditions for the pseudo-linear
approximation, but would be violated, in the absence of non-linear
effects, by those that satisfy the thin-shell conditions.  We must
therefore conclude that non-linear effects near the surface of a
body with thin-shell ensure $\phi > \phi_{crit}$ is always
satisfied as $r \rightarrow \infty$. Furthermore, if $\phi \gg
\phi_{crit}$ then
$$
\left\vert \frac{d \phi}{d r} \right\vert \ll \frac{\phi_b
R}{r^2}
$$
and it follows from Section \ref{singpseu} that the pseudo-linear
approximation is valid for all $r$, which further implies that the
body {cannot} have a thin-shell.  We are therefore justified
in using $\phi_{crit}$ to approximate the far field of a body with
a thin-shell.

There are a number of consequences of the existence of a critical form for $\phi$ when $r \gg R$. \mbox{In particular,}
 no matter how massive our central body is, no matter how
strongly it is coupled to the chameleon, the perturbation it produces
in $\phi$ for $r \gg R$ takes a universal value whenever the
thin-shell conditions are valid.

Remarkably, the far field is found to be {independent} of the
coupling, $\beta$, of the chameleon to the isolated body.
The $\beta$ independence
is a generic feature of all $V \sim \phi^{-n}$ theories (\emph{i.e.}, not
only for \mbox{$n>0$ theories} but also for $n<-4$ and $n=-4$). The far field of a body with a
thin-shell is independent of $\beta$, and so, in contrast to
what occurs for linear theories, larger values of $\beta$ do
\emph{not} result in larger forces between distant bodies.

Let us define the mass of the body as ${\mathcal{M}}=4\pi
\rho_c R^3/3$. Then we can express this critical behaviour of the far
field in terms of an effective coupling, $\beta_{eff}$, defined
by:
$$ \phi \sim \phi_b - \frac{\beta_{eff} {\mathcal{M}}e^{-m_b r}}{4\pi
M_{pl}r}
$$ when $r \gg R$.  If we assume $\rho_{b}/\rho_c \ll 1$, we have (see \cite{Mota:2006fz}):
\begin{eqnarray}
\beta_{eff} &=& \frac{4\pi
M_{pl}}{{\mathcal{M}}}MR\left(\frac{n(n+1)M^2}{m_b^2}\right)^{\frac{1}{n+2}},
\qquad n > 0 \label{critbeta}
\end{eqnarray}

This $\beta$-independence means that if one uses test-bodies with
the same mass and outer dimensions, then in chameleon theories, no
matter how much the weak equivalence principle is violated at a
particle level, there will be \emph{no} violations of weak equivalence principle (WEP) far from
the body, because the far field is totally independent of both the
body's chameleon coupling and its density. In other words, the
chameleon can be coupled to neutrons and protons in different ways
(\emph{i.e.}, $\beta^{(n)} \neq \beta^{(p)}$) but no violation of WEP will
be detected (\emph{i.e.}, $\beta^{(n)}_{eff}=\beta^{(p)}_{eff}$).
Moreover, as the reader can see from Equation (\ref{critbeta}), also
$M_p$ and ${\mathcal{M}}$ are cancelled from the chameleon when
the thin-shell mechanism is present. This remarkable and
surprising cancellation is a major difference between chameleonic
and standard gravity.

\subsubsection{$m_c R>>1$ and the Harmonic Approximation}
\label{armonica}

Following \cite{Waterhouse:2006wv} we now assume that outside the
sphere $V_\text{eff}(\phi)$ can be approximated
by a harmonic potential. Then for $r>R$,
\begin{displaymath}
\frac{d^2\phi}{dr^2}+\frac{2}{r}\frac{d\phi}{dr}=m_b^2\left(\phi-\phi_b\right)
\end{displaymath}

The general solution to this differential equation is
\begin{equation}
\phi(r)=A\frac{e^{-m_b(r-R)}}{r}+B\frac{e^{m_b(r-R)}}{r}+\phi_b
\label{harmonicsolution}
\end{equation}
for dimensionless constants $A$ and $B$.  The condition
 $\phi\rightarrow\phi_b$ as $r\rightarrow\infty$ gives $B=0$
so we have
\begin{equation}
\phi(r)=A\frac{e^{-m_b(r-R)}}{r}+\phi_b \label{outsidesolution}
\end{equation}

In \cite{Waterhouse:2006wv} the author for $r<R$ considered two
classes of solution based on two different approximations.
Firstly, the author defined $R_\text{c}$ to divide the interval
$\left[0,R\right]$ into two intervals: $\left[0,R_\text{c}\right]$
on which $\phi\sim\phi_\text{c}$, and $\left[R_\text{c},R\right]$
on which $\phi\gg\phi_\text{c}$. Secondly, the chameleonic
equation has been solved in each interval, following two different
approximations:
\begin{description}
\item[Approximation 1:]  $\phi\gg\phi_\text{c}$. In this case the
harmonic approximation to $V_\text{eff}$ is not valid, but the
bare potential $V$ decays quickly and the term $\rho
e^{\beta\phi/M_\text{pl}}$ comes to dominate. In particular, we
have (for both the power-law potential and the exponential
potential)
\begin{equation}
V_{\text{eff},\phi}\left(\phi\right)\approx\frac{\beta}{M_\text{pl}}\rho_\text{c}
\label{linearapproximation}
\end{equation}

The equation for the chameleon now takes the form
\begin{displaymath}
\frac{d^2\phi}{dr^2}+\frac{2}{r}\frac{d\phi}{dr}\approx\frac{\beta}{M_\text{pl}}\rho_\text{c}
\end{displaymath}
with the general solution
\begin{equation}
\phi\left(r\right)=\frac{\beta}{6M_\text{pl}}\rho_\text{c}r^2+\frac{C}{r}+D\phi_\text{c}
\label{inharmonicsolution}
\end{equation}
for dimensionless constants $C$ and $D$. \item[Approximation 2:]
$\phi\sim\phi_\text{c}$. We can use the harmonic approximation
like we did in Equation (\ref{harmonicsolution}):
\begin{equation}
V_{\text{eff},\phi}\left(\phi\right)\approx
m_\text{c}^2\left(\phi-\phi_\text{c}\right),
\label{harmonicapproximation}
\end{equation}
but this time we will write the solution as
\begin{equation}
\phi(r)=E\frac{e^{-m_\text{c}r}}{r}+F\frac{e^{m_\text{c}\left(r-R_\text{c}\right)}}{r}+\phi_\text{c},
\label{otherharmonicsolution}
\end{equation}
where $E$ and $F$ are, of course, dimensionless constants.
\end{description}

By imposing the continuity conditions and the requirement
$\frac{d\phi}{dr}\rightarrow0$ as $r\rightarrow0$ to ensure
continuity of the three-dimensional solution at the origin, we
obtain an approximate (thin-shell) solution.

Since $E=-Fe^{-m_\text{c}R_\text{c}}$, the solution has been given
in \cite{Waterhouse:2006wv} in the form:
\begin{gather*}
\phi(r)=\left\{
\begin{aligned}
&F\frac{e^{m_\text{c}\left(r-R_\text{c}\right)}-e^{-m_\text{c}\left(r+R_\text{c}\right)}}{r}+\phi_\text{c} & r&\in\left(0,R_\text{c}\right)\\
&\frac{\beta}{6M_\text{pl}}\rho_\text{c}r^2+\frac{C}{r}+D\phi_\text{c} & r&\in\left(R_\text{c},R\right)\\
&A\frac{e^{-m_\infty(r-R)}}{r}+\phi_\infty &
r&\in\left(R,\infty\right)
\end{aligned}
\right.
\end{gather*}

In \cite{Waterhouse:2006wv}, the discussion of the thin-shell
solution started from the assumption $F=0$, namely $\phi \simeq
\phi_c$ inside the body (following Khoury and Weltman). The next
step in \cite{Waterhouse:2006wv} was to show, with the help of the
continuity equations, that the chameleon field far from the body
takes the form:

\begin{equation}
\phi_\text{thin}\left(r\right)\approx-\frac{\beta}{4\pi
M_\text{pl}}\left(\frac{4}{3}\pi
R^3\rho_\text{c}\right)\left[3\frac{M_\text{pl}
\left(\phi_\infty-\phi_\text{c}\right)}{\beta\rho_\text{c}R^2}\right]\frac{e^{-m_\infty(r-R)}}{r}+\phi_\infty
\label{thinfattore}
\end{equation}
where the factor in the square brackets is present only in the
thin-shell case. It is precisely this factor that implies the
$\beta$, $M_p$ and ${\mathcal{M}}$-independence mentioned above
(see {Equation} (\ref{critbeta})).

Now we would like to add some comments. Firstly, we point out that
there exists a choice of parameters such that the chameleonic
solution for $R_c<r<R$ is basically indistinguishable from
$\phi_c$. Indeed we have:

\begin{equation}
\frac{\phi}{\phi_c}=1+\frac{\beta}{M_p} \frac{\rho_c
R_c^2}{\phi_c} [ \frac{1}{6} (\frac{r}{R_c})^2 +\frac{R_c}{3 r}-\frac{1}{2}]
\label{formuletta}
\end{equation}

If we define $\epsilon=\frac{\Delta R}{R_c}$ and we evaluate the
square bracket in the limit $\epsilon\rightarrow0$ (\textit{i.e}
thin-shell), we find that the factor in the square bracket is
negligible. As far as the prefactor $\frac{\beta}{M_p}
\frac{\rho_c R_c^2}{\phi_c}$ is concerned, its value will be
determined by the choice of the parameters of the model.
Remarkably, there are models where $\phi \simeq \phi_c$ in the
shell, for example the models where the mass scale in the scalar potential is not fine-tuned and, therefore, it is much larger than the meV-scale. For this reason we will now consider a 2-regions set-up,
namely we will exploit an harmonic approximation everywhere inside
the body with thin shell.

The chameleon field in this approximation can be written as:
\begin{gather*}
\phi(r)=\left\{
\begin{aligned}
&\tilde{F} \frac{e^{m_\text{c}\left(r-R\right)}-e^{-m_\text{c}\left(r+R\right)}}{r}+\phi_\text{c} & r&\in\left(0,R\right)\\
&A\frac{e^{-m_\infty(r-R)}}{r}+\phi_\infty
&r&\in\left(R,\infty\right)
\end{aligned}
\right.
\end{gather*}
where we introduced a new constant $F e^{-m_c R_c}\rightarrow
\tilde{F} e^{-m_c R}$.

If we impose the continuity conditions at the surface for the
chameleon and its derivative we obtain
\begin{equation}
\frac{\tilde{F}}{R} (1-e^{-2 m_c R}) +\phi_c=\phi_b +\frac{A}{R}
\end{equation}
from the chameleon and
\begin{equation}
\tilde{F}= [m_c R (1+ e^{-2 m_c R})+e^{-2 m_c R }-1]=-A(1+m_b R)
\end{equation}
from the first derivative of the chameleon.

Now we consider the thin-shell case: we assume $m_c R>>1$ and we
neglect the exponential terms like $e^{-m_cR}$. In this way we
obtain:

\begin{equation}
\tilde{F}= (\phi_b-\phi_c) \frac{1+m_b R}{m_c +m_b}\\
A=(\phi_b-\phi_c) \frac{1-m_cR}{m_c+m_b}
\end{equation}

Consequently we can write the (thin-shell) chameleon field as:

\begin{gather*}
\phi(r)=\left\{
\begin{aligned}
&(\phi_b-\phi_c) \frac{1+m_b R}{m_c +m_b} \frac{e^{m_c (r-R)}-
e^{-m_c (r+R)}}{r} +\phi_c  & r&\in\left(0,R\right)   \\
& \phi_b + (\phi_b-\phi_c) \frac{1-m_cR}{m_c+m_b} \frac{e^{-m_b
(r-R)}}{r} & r&\in\left(R,\infty\right)
\end{aligned}
\right.
\end{gather*}

Remarkably, if we exploit the thin-shell condition $m_c R>>1$ once
again and we write $\frac{1-m_cR}{m_c+m_b} \rightarrow -R$, we
obtain the same solution already mentioned in Equation (\ref{thinfattore}).
The $\beta$, $M_p$ and ${\mathcal{M}}$-independence is recovered
once again. If we focus our attention on the external (thin-shell)
solution close to the body, we have $\phi \simeq \phi_c$. There
are a number of consequences of this fact:
\begin{itemize}
\item
The gravitational potential of the theory will {not} depend on
${\mathcal{M}}$ like in a standard gravitational theory. This
remarkable independence of the solution on the mass of the object
can be connected to the breakdown of the superposition principle
in the non-linear theory we are considering. \vspace{-9pt}
\item The non-linear
nature of the chameleonic theory close to the body cannot be
neglected. Since $\phi_b$ minimizes the effective potential close
to the body and we have $\phi\simeq \phi_c\neq\phi_b$, then the
harmonic approximation will not be valid and a more complete
analysis is necessary following Section \ref{nonclose}.
\end{itemize}

\subsubsection{The Chameleon Force} The interaction of the
chameleon field with matter is encoded in the conformal coupling
of \mbox{Equation (\ref{conformalrelation}).} Since matter fields
$\psi_\text{m}^{(i)}$ couple to $g_{\mu\nu}^{(i)}$ instead of
$g_{\mu\nu}$, the worldlines of free test particles (meaning
particles experiencing only gravity and the chameleon force) of
species $i$ are the geodesics of $g_{\mu\nu}^{(i)}$ rather than
those of $g_{\mu\nu}$.

The geodesic equation for the worldline $x^\mu$ of a test mass of
species $i$ is
\begin{equation}
\ddot{x}^\rho+\tilde{\Gamma}_{\mu\nu}^{\rho}\dot{x}^\mu\dot{x}^\nu=0
\label{geodesic}
\end{equation}
where $\tilde{\Gamma}_{\mu\nu}^{\rho}$ are the Christoffel symbols
and a dot denotes differentiation with respect to proper time
$\tilde{\tau}$, both in the $\tilde{g}_{\mu\nu}$ metric.

Using
\begin{displaymath}
\tilde{g}_{\mu\nu,\sigma}=\left(\frac{2\beta_i}{M_\text{pl}}\phi_{,\sigma}g_{\mu\nu}+g_{\mu\nu,\sigma}\right)e^{2\beta_i\phi/M_\text{pl}}
\end{displaymath}
the Christoffel symbols can be determined as follows:
\begin{align*}
\tilde{\Gamma}_{\mu\nu}^{\rho}&=\frac{1}{2}\tilde{g}^{\sigma\rho}\left(\tilde{g}_{\sigma\nu,\mu}+\tilde{g}_{\sigma\mu,\nu}-\tilde{g}_{\mu\nu,\sigma}\right)\\
&=\Gamma_{\mu\nu}^\rho+\frac{\beta_i}{M_\text{pl}}\left(\phi_{,\mu}\delta_\nu^\rho+\phi_{,\nu}\delta_\mu^\rho-g^{\sigma\rho}\phi_{,\sigma}g_{\mu\nu}\right)
\end{align*}

Substituting this into Equation (\ref{geodesic}) gives
\begin{align*}
0&=\ddot{x}^\rho+\Gamma_{\mu\nu}^{\rho}\dot{x}^\mu\dot{x}^\nu+\frac{\beta_i}{M_\text{pl}}\left(2\phi_{,\mu}\dot{x}^\mu\dot{x}^\rho+g^{\sigma\rho}\phi_{,\sigma}\right)
\end{align*}

The second term in the above equation is the familiar
gravitational term, while the term in $\beta_i/M_\text{pl}$ is the
chameleon force.

We see that in the non-relativistic limit, a test mass $m$ of
species $i$ in a static chameleon field $\phi$ experiences a force
$\vec{F}_\phi$ given by
\begin{equation}
\frac{\vec{F}_\phi}{m}=-\frac{\beta_i}{M_\text{pl}}\vec{\nabla}\phi
\label{chameleonforce}
\end{equation}
as in \cite{Khoury:2003aq,Khoury:2003rn}.  Thus, $\phi$ is
the potential for the chameleon force.

We will now consider the force between two bodies, with
thin-shells, that are separated by a distance $d \gg R_1, \, R_2$,
where $R_1$ and $R_2$ are respectively the length scales of body
one and body two.

We expect that, outside some thin region close to the surface of
either body, the pseudo-linear approximation is appropriate to
describe the field of either body. In the region where
pseudo-linear behaviour is seen, we can safely super-impose the
two 1-body solutions to find the full 2-bodies solution.

The perturbation to $\phi$ induced by body two near body one will
be $$ \delta \phi_{2} = -\frac{\beta_{eff}^{(2)} {\mathcal{M}}_2
e^{-m_b d}}{4 \pi M_{pl}d} $$

From the results of the previous
{(sub)section}, we know that
$\beta_{eff}^{(2)}{\mathcal{M}}_{2}/M_{pl}$ depends only on the
radius of body two and on the theory-dependent parameters, $M$,
$\lambda$ and $n$ (see Equation (\ref{critbeta})). In our usual
notations, $m_b$ is the mass of the chameleon field in the
background and ${\mathcal{M}}_{2}$ is the mass of body two.

The perturbation to $\phi$ induced by body one near body two is
$$ \delta \phi_{1} = -\frac{\beta_{eff}^{(1)} {\mathcal{M}}_1
e^{-m_b d}}{4 \pi M_{pl}d} $$

From Equation (\ref{critbeta}) we know
that $\delta \phi_{1}$ is independent of $\beta$ and the mass of
body one, ${\mathcal{M}}_{1}$. The force on body one due to body
two will be proportional to $\nabla \delta \phi_{2}$, however,
since this must also be the force on body two due to body one, it
must also be proportional to $\nabla \delta \phi_{1}$ evaluated
near body two. Consequently, the force on one body due to the
other must, up to a possible ${\mathcal{O}}(1)$ factor, be given
by \cite{Mota:2006fz}: $$ F_{\phi} =
\frac{\beta_{eff}^{(1)}{\mathcal{M}}_{1}\beta_{eff}^{(2)}{\mathcal{M}}_{2}
(1+m_b d)e^{-m_b d}}{(4\pi M_{pl})^2 d^2} $$

The functional
dependence of this force on $M$, $n$, $R_1$, $R_2$ and $\lambda$
depends on whether \mbox{$n < -4$,}\mbox{ $n=-4$ or} $n >0$. In all cases the
force is found to be independent of $\beta$, ${\mathcal{M}}_1$ and
${\mathcal{M}}_2$. For a more detailed discussion see
\cite{Mota:2006fz}.

It seems noteworthy considering the case where one of the two
bodies does {not} have a thin shell \cite{Mota:2006fz}. In
this case the $\beta$-independence is lost (see also Equation (\ref{chameleonforce})):

$$ F_{\phi}
= \frac{\beta_{eff}^{(1)}\beta {\mathcal{M}}_1
{\mathcal{M}}_2(1+m_bd)e^{-m_bd}}{4\pi d^2} $$ whenever $d \gg
R_1$ where $R_1$ is the radius of curvature of body one and
$\beta_{eff}^{(1)}$ is given by Equation (\ref{critbeta}) for
potentials with $n>0$.

\section{Chameleonic Gravity}

Chameleon fields can be a very useful guideline towards the description of alternative theories of gravity. In this section we will further explore this issue. In particular, we will briefly analyze the connection between chameleon fields, {$f(R)$} theories and quantum gravity. For more details the reader is referred to \cite{Zanzi:2014twa, Capozziello:2011et, Capozziello:2007eu} and related references.

\subsection{f(R) Gravity and Chameleon Fields}

$f(R)$ theories of gravity are characterized by a modified gravitational lagrangian where, instead of the standard curvature $R$, we find a general function of the Ricci scalar. In these theories, some stability conditions must be fulfilled in order to avoid the presence of tachyons and ghosts for $R\geq R_1$, where $R_1$ is a de Sitter point. In particular we must have $f_{,R}>0$ and $f_{,RR}>0$ for a stable model. Interestingly, some $f(R)$ models can fulfill the stability conditions and also provide the correct cosmological sequence of radiation, matter and accelerated phase. In these models, solar system constraints can be faced with the help of the chameleon mechanism. Let us further illustrate this point.

Let us consider the action \cite{Capozziello:2007eu}:
\begin{equation}
\label{action}
S=\int{\rm d}^{4}x\sqrt{-g}\,
f(R)/2+S_{m}(g_{\mu \nu}, \Psi_m)\,
\end{equation}
where $S_m$ is a matter Lagrangian
that depends on the metric $g_{\mu \nu}$ and on matter
fields $\Psi_m$. We have chosen $M_{\rm pl}^2=(8\pi G)^{-1}=1$.

We can perform a conformal transformation introducing a new metric $\tilde{g}_{\mu \nu}$
and a scalar field $\phi$, as
\begin{eqnarray}
\tilde{g}_{\mu \nu}=\psi g_{\mu \nu}\,, \quad
\phi=\sqrt{3/2}\,{\rm ln}\,\psi\,
\end{eqnarray}
where $\psi=\partial f/\partial R$.
Then the action in the E-frame is
\begin{eqnarray}
\label{actione}
S &=&\int{\rm d}^{4}x\sqrt{-\tilde{g}}\left[
\tilde{R}/2-(\tilde{\nabla}\phi)^2/2
-V(\phi) \right] \nonumber \\
& &+S_m (\tilde{g}_{\mu \nu} e^{2\beta \phi}, \Psi_m)\,
\end{eqnarray}
where
\begin{eqnarray}
\beta=-\frac{1}{\sqrt{6}}\,,\quad
V=\frac{R(\psi)\psi-f}{2\psi^2}\,.
\end{eqnarray}

The field $\phi$ is directly coupled to
matter with a constant coupling $\beta$ and it is a chameleon field.

Let us study some examples of $f(R)$ models that satisfy, on the one hand,
local gravity constraints, on the other hand, cosmological and stability
conditions. There is the Hu-Sawicki model
\begin{eqnarray}
\label{mo1}
f(R)=R-\lambda R_c \frac{(R/R_c)^{2n}}{(R/R_c)^{2n}+1}\,
\end{eqnarray}
and the Starobinsky one
\begin{eqnarray}
\label{mo2}
f(R)=R-\lambda R_c \left[ 1-\left( 1+R^2/R_c^2
\right)^{-n} \right]\,
\end{eqnarray}

In both models $n$, $\lambda$ and $R_c$ are positive constants.
These two models share a common $f$ function in the large curvature limit $R>>R_c$.

Let us discuss post-Newtonian solar-system constraints on
these two models in the large curvature regime.
In the weak-field approximation the rotationally invariant
metric in the J-frame is

\newpage
\vspace{-12pt}
\begin{eqnarray}
{\rm d}s^2=-[1-2{\cal A}(r)]{\rm d}t^2
+[1+2{\cal B}(r)]{\rm d}r^2+r^2 {\rm d}
\Omega^2\,
\end{eqnarray}
where ${\cal A}(r)$ and ${\cal B}(r)$ are the functions
of $r$. The post-Newtonian parameter,
$\gamma={\cal B}(r)/{\cal A}(r)$,
is roughly given by \cite{Faul}
\begin{eqnarray}
\gamma \simeq \frac{1-\Delta r_c/r_c}
{1+\Delta r_c/r_c}\,
\end{eqnarray}

In this formula, when we wrote the thin-shell parameter we used a lower case $r$ in order to avoid confusion with the scalar curvature $R$.
The present tightest constraint on $\gamma$ is
$|\gamma-1| <2.3 \times 10^{-5}$ which
implies a bound on the thin-shell parameter.

For the large curvature model the de Sitter point corresponds to
$\lambda=x_1^{2n+1}/(2(x_1^{2n}-n-1))$, where
$x_1=R_1/R_c$ (see \cite{Capozziello:2007eu}).
Consequently, the bound induced by $\gamma-1$ gives
\begin{eqnarray}
\label{cons2}
\frac{n}{2(x_1^{2n}-n-1)}\left(
\frac{R_1}{\rho_B} \right)^{2n+1}<2.4 \times 10^{-11}\,
\end{eqnarray}

For the stability of the de Sitter point we must have (see \cite{Capozziello:2007eu}) $x_1^{2n}>2n^2+3n+1$.
Hence the term $n/2(x_1^{2n}-n-1)$ in Equation~(\ref{cons2})
is smaller than 0.25 for $n>0$.
If we assume $R_1$ of the order of the present
cosmological density $10^{-29}$ g/cm$^3$ and $\rho_B$ roughly comparable to the
baryonic/dark matter density in the Milky Way ($10^{-24}$ g/cm$^3$), we are led to
\begin{eqnarray}
\label{cons3}
n>0.5\,
\end{eqnarray}
\subsection{Chameleonic Quantum Gravity}

In the previous paragraphs we described gravity through a metric (and a scalar field). A metric can be safely exploited granted that a large number of gravitons is assumed. Naturally, we are free to perform quantum loops and we would obtain a semi-classical description of gravity, which is not yet quantum gravity (QG). In order to talk about a QG model, two conditions must be fulfilled: (1) we must take into account quantum effects (\emph{i.e.}, $\hbar$ is non-vanishing); and (2) we must have a finite number $N$ of gravitons (\emph{i.e.}, we avoid the limit $N \rightarrow \infty $).

Remarkably, chameleon fields can be a useful guideline towards QG. In a recent paper \cite{Zanzi:2014twa} a Chameleonic Equivalence Principle (CEP) has been discussed in the framework of the so-called Modified Fujii's Model (MFM). Here is the model. We have two different conformal frames: (1) the string frame (S-frame) where the cosmological constant (CC) is large (for example planckian) and the fields are stabilized (including the dilaton $\phi$); (2) the E-frame with a chameleonic dilaton $\sigma$ parametrizing the amount of scale invariance. This scale symmetry is abundantly broken locally (\emph{i.e.}, ``in this room'') and it is almost restored globally (\emph{i.e.}, on cosmological distances) in the E-frame. Consequently,  the (renormalized) vacuum energy in the E-frame is running from large values in the UV to small values in the IR.

We write the string frame lagrangian as (the gauge part is not written explicitly but it is present in \mbox{the theory)}
\beq {\cal L}={\cal L}_{SI} + {\cal L}_{SB}, \label{Ltotale}\eeq where the
Scale-Invariant part of the Lagrangian is given by:

\begin{equation}
{\cal L}_{\rm SI}=\sqrt{-g}\left( \half \xi\phi^2 R -
    \half\epsilon g^{\mu\nu}\partial_{\mu}\phi\partial_{\nu}\phi -\half g^{\mu\nu}\partial_\mu\Phi \partial_\nu\Phi
    - \frac{1}{4} f \phi^2\Phi^2 - \frac{\lambda_{\Phi}}{4!} \Phi^4
    \right)
\label{bsl1-96}
\end{equation}
$R$ is the curvature. $\Phi$ is a scalar field representative of matter fields,
$\epsilon=-1$, $\left( 6+\epsilon\xi^{-1} \right)\equiv
\zeta^{-2}\simeq 1$, $f<0$ and $\lambda_{\Phi}>0$.
We can also write terms of the form $\phi^3 \Phi$, $\phi \Phi^3$ and
$\phi^4$ which are scale invariant, however,
we will not include these terms. Indeed, the first two terms can be removed exploiting symmetries of strong interaction {\cite{Fujii}} and the $\phi^4$ term does not clash with the solution to the CC problem, because the {renormalized Planck mass in the IR region is an exponentially decreasing function of $\sigma$ (see also \cite{Zanzi:2012du}). In this} {model the Planck mass is not constant and, interestingly, it is related to the masses of particles in the E-frame.}

The Symmetry Breaking Lagrangian
${\cal L_{SB}}$   contains scale-non-invariant terms,
in particular, a stabilizing potential for $\phi$ in the
S-frame. For this reason we write: \beq {\cal L}_{\rm
SB}=-\sqrt{-g} (a \phi^2 + b + c \frac{1}{\phi^2}) \label{SB}
\eeq

The reader can find a possible choice of parameters in \cite{Zanzi:2012du}. For a discussion of the E-frame lagrangian see \cite{Zanzi:2010rs}.

As pointed out in {\cite{Zanzi:2014twa}}, the MFM satisfies a Chameleonic Equivalence Postulate (CEP).
Here is the CEP:
{\it {for each pair} of vacua V1 and V2 allowed by the theory there is a conformal transformation that connects them and such that the mass of matter fields $m_0,_{V1}$ ({i.e.}, $m_0$ evaluated in V1) is mapped to $m_0,_{V2}$ ({i.e.}, $m_0$ evaluated in V2). When a conformal transformation connects two vacua with a different amount of conformal symmetry, an additional term (in the form of a conformal anomaly) must be included in the field equations and this additional term is equivalent to the gravitational field.}

The CEP is a consequence of the chameleon mechanism and, in this sense, it is a principle, not a postulate.
We learn that the chameleon mechanism is suggesting us an equivalence between the quantum gravitational field and a conformal anomaly. Let us discuss some gravitational aspects of the CEP.

Special Relativity (SR) is based on an invariance principle: the laws of Nature are invariant under Lorentz transformations. Needless to say, SR is not a theory of gravity. If our intention is to describe gravity in a relativistic way, one possibility is to consider GR which is based on the Equivalence Principle (EP). The EP is telling us that gravitation is equivalent to inertia. It is common knowledge that whenever we perform a transformation that connects an inertial frame to a non-inertial one, additional terms will be present in the equations of the theory. For example we can consider the Newton equations: in non-inertial frames the Newton equations acquire some additional terms (the inertial forces). This idea is valid
also in GR. When we perform a general coordinate transformation in GR, some ``additional terms'' (the metric and the connection) will be present in the equations of the theory and these terms are exploited by Einstein to describe the gravitational field in harmony with the EP. GR is a classical theory of gravity.

Now we move to QG. How is it possible to describe a complete absence of gravity? Our GR-based intuition tells us that we must remove all the masses/energy sources (including vacuum energy!). In the E-frame, the low-redshift cosmological vacuum of the MFM has basically no gravity. If our intention is to switch on a (small) gravitational field we can add a source given by, for example, a massive (or even massless) particle and the amount of conformal symmetry will be slightly reduced by the chameleon mechanism. In this way, whenever we modify the gravitational field, we obtain a chameleonic shift of the ground state and, see \cite{Zanzi:2014twa} and references therein,  this jump can be summarized by a conformal anomaly. In other words, in the MFM, the chameleon mechanism is telling us that the quantum gravitational field is described by a conformal anomaly in harmony with the CEP. The CEP is a guideline towards the QG regime. In particular, let us construct the following dictionary for QG:
\begin{itemize}
\item Let us replace the inertial frame of Einstein's theories with a conformal ground state in the MFM. Let us write the connection between the two models in this way: inertial frame $ \rightarrow$ conformal ground state. \vspace{-9pt}
\item Non-inertial frame $\rightarrow$ non-conformal ground state.\vspace{-9pt}
\item General coordinate transformation $\rightarrow$ conformal transformation.\vspace{-9pt}
\item Metric and connection $\rightarrow$ conformal anomaly.
\end{itemize}

In this way, the ``dictionary'' mentioned above creates a connection between classical and QG. \mbox{The CEP} is the microscopic counterpart of the EP.

\section{Conclusions}

Chameleon fields are particles with an environment dependent mass. When the matter density is large, the field is stabilized. When the matter density is small (for example on cosmological distances) the field is light. Hence the name chameleon. This peculiar behaviour of the mass is particularly interesting for phenomenological reasons. Indeed, it is common knowledge that it is not possible to build a model where a very light scalar degree of freedom has a generic coupling to matter, because phenomenological constraints must be faced. The chameleon mechanism (and the related thin-shell) can screen the scalar field locally. However, there are also theoretical reasons to consider chameleon fields. For example they are a useful guideline towards a theory of quantum gravitation (as we saw when we discussed the CEP). One of the remarkable properties of chameleonic models is that they are expected to be testable or falsifiable by experiments which can be alternative to accelerator-based experiments. Needless to say, accelerators are the ``main road'' to search for physics beyond the SM, however, it is good to dedicate attention to alternative paths in our search for experimental signatures of high energy physics theories. Chameleon fields are in harmony with this idea.


\acknowledgments{Acknowledgments}

Special thanks are due to Antonio Masiero and Massimo Pietroni for many enlightening discussions about chameleon fields. I warmly thank the journal {{\it Universe}} and its staff for inviting me to contribute to this special issue. I also thank two reviewers of the journal for very useful comments on a preliminary edition of the manuscript.
%

\bibliographystyle{mdpi}

\begin{thebibliography}{----} 


\bibitem{Kazakov:2006kp}
Kazakov, D.; Lavignac, S.; Dalibard, J.  \textit{Particle Physics
Beyond the Standard Model}; Elsevier: Amsterdam, The Netherland, 2006.


\bibitem{Luest:2013ll}
Blumenhagen, R.; Luest, D.; Theisen, S.  {\em Basic Concepts of String
Theory}; {Springer-Verlag: {Berlin}, {Germany}}, 2013; pp. 1--782.

\bibitem{Correia:2007sv}
Correia, F.P.; Schmidt, M.G.A.   Moduli stabilization in heterotic
  M-theory. {\em  Nucl. Phys. B} \textbf{2008}, {\em 797},  243--267.

\bibitem{Khoury:2003aq}
Khoury, J.; Weltman, A. Chameleon fields: Awaiting surprises for tests of
  gravity in space.
\mbox{{\em Phys. Rev. Lett.}} \textbf{2004},
  {\em 93}, {171104}.



\bibitem{Khoury:2003rn}
Khoury, J.; Weltman, A. Chameleon cosmology.
  {{\em Phys. Rev. D} \textbf{2004}, {\em
  69},  {044026}}.


\bibitem{Zanzi:2010rs}
Zanzi, A.  {Chameleonic dilaton, nonequivalent frames, and the cosmological
  constant problem in quantum string theory}.
  {{\em Phys. Rev. D} \textbf{2010}, {\em
  82},  {044006}}.

\bibitem{Zanzi:2012du}
Zanzi, A. {\em Chameleonic Dilaton and Conformal Transformations}; The Cornell University:  Ithaca, NY, USA, 2012.

\bibitem{Zanzi:2012ha}
Zanzi, A. {\em Species, Chameleonic Strings and the Concept of Particle}; The Cornell University:  Ithaca, NY, USA, 2012.

\bibitem{Zanzi:2012bf}
Zanzi, A. {\em Dilaton Stabilization and Composite Dark Matter in the String
  Frame of Heterotic-M-Theory}; The Cornell University:  Ithaca, NY, USA, 2012.

\bibitem{Zanzi:2014twa}
Zanzi, A. {Chameleonic Equivalence Postulate and Wave Function Collapse}.
  {\em Electron. J. Theor. Phys.} \textbf{2015}, {\em 12},  1--28.

\bibitem{Vainshtein:1972sx}
Vainshtein, A.  {To the problem of nonvanishing gravitation mass}. {{\em Phys. Lett. B}  \textbf{1972}, {\em
  39},  393--394}.

\bibitem{Hinterbichler:2010es}
Hinterbichler, K.; Khoury, J. {Symmetron Fields: Screening Long-Range
  Forces Through Local Symmetry Restoration}.
  {{\em Phys. Rev. Lett.} \textbf{2010},
  {\em 104}, 231301}.

\bibitem{Pietroni:2005pv}
Pietroni, M.  {Dark energy condensation}.
  {{\em Phys. Rev. D} \textbf{2005}, {\em 72},  043535}.

\bibitem{Olive:2007aj}
Olive, K.A.; Pospelov, M.   {Environmental dependence of masses and coupling
  constants}.
   {\em
 \mbox{ Phys. Rev. D}} \textbf{2008}, {\em 77},  043524.

\bibitem{Hinterbichler:2010wu}
Hinterbichler, K.; Khoury, J.; Nastase, H. {Towards a UV Completion for
  Chameleon Scalar Theories}.
  {{\em J. High. Energ. Phys.} \textbf{2011}, {\em 1103}, 061}.

\bibitem{Nastase:2013ik}
Nastase, H.; Weltman, A.  {Chameleons on the Racetrack}.
  {{\em J. High. Energ. Phys.} \textbf{2013}, {\em 1308},
   059}.

\bibitem{Brax:2011qs}
Brax, P.;  Davis, A.-C.  {Supersymmetron}.
  {{\em Phys. Lett. B} \textbf{2012}, {\em
  707},  1--7}.

\bibitem{Brax:2012mq}
Brax, P.;  Davis, A.-C.; Sakstein, J.  {SUPER-Screening}.
  {{\em Phys. Lett. B} \textbf{2013},
  {\em 719},  210--217}.

\bibitem{Hinterbichler:2013we}
Hinterbichler, K.; Khoury, J.; Nastase, H.; Rosenfeld, R.  {Chameleonic
  inflation}.
\textbf{   \href{http://dx.doi.org/10.1007/JHEP08(2013)053}}
   {{\em J. High. Energ. Phys.} \textbf{2013},
  {\em 08},~053}.

\bibitem{Adelberger:2006dh}
Adelberger, E.; Heckel, B.R.; Hoedl, S.A.; Hoyle, C.; Kapner, D.; Upadhye, A.
   {Particle Physics Implications of a Recent Test of the Gravitational
  Inverse Sqaure Law}.
  {{\em Phys. Rev. Lett.} \textbf{2007},
  {\em 98},~131104}.

\bibitem{Steffen:2010ze}
Steffen, J.H.; Upadhye, A.;  Baumbaugh, A.;  Chou, A.S.; Mazur, P.O.; Tomlin, R.;  \mbox{Weltman, A.;}  \mbox{Wester, W.} {Laboratory constraints on chameleon dark energy
  and power-law fields}. \emph{\mbox{Phys. Rev. Lett.}} \textbf{2010}, \emph{105}, 261803.
%

\bibitem{Rybka:2010ah}
Rybka, G.; Hotz, M.; Rosenberg, L.J.; Asztalos, S.J.; Carosi, G.; Hagmann, C.; Kinion, D.; \mbox{van Bibber, K.;} Hoskins, J.; Martin, C.;  {\em et al}.  {A Search for Scalar Chameleons with ADMX}.
  {{\em Phys. Rev.
  Lett.} \textbf{2010}, {\em 105},  051801}.
\newpage
\bibitem{Hu:2013aqa}
  Hu, B.; Liguori, M.; Bartolo, N.; Matarrese, S.
   {Parametrized modified gravity constraints after Planck}.
  {{\em Phys. Rev.
  D} \textbf{2013}, {\em 88}, 123514}.

\bibitem{Brax:2010xq}
Brax, P.; Zioutas, K. {Solar Chameleons}.
\emph{Phys. Rev. D} \textbf{2010}, \emph{82}, 043007.


\bibitem{Davis:2009vk}
Davis, A.-C.;  Schelpe, C.A.O.; Shaw, D.J.   {The Effect of a Chameleon
  Scalar Field on the Cosmic Microwave Background}.
  {{\em Phys. Rev. D} \textbf{2009}, {\em
  80}, 064016}.

\bibitem{Davis:2010nj}
Davis, A.-C.;  Schelpe, C.A.O.; Shaw, D.J.   {The Chameleonic Contribution
  to the SZ Radial Profile of the Coma Cluster}. \emph{Phys. Rev. D} \textbf{2011}, \emph{83}, 044006.


\bibitem{Khoury:2013yy}
Khoury, J.  {Chameleon Field Theories}. \mbox{\emph{Classical Quant. Grav.} \textbf{2013},}
doi:10.1088/0264-9\linebreak 381/30/21/214004.


\bibitem{Zanzi:2014aia}
Zanzi, A.; Ricci, B.  {Chameleon fields and solar physics}.
  {{\em Mod. Phys. Lett. A} \textbf{2015},
  {\em 30}, 1550053}.

\bibitem{Zanzi:2015kha}
Zanzi, A.
   {Chameleon fields, wave function collapse and quantum gravity}.
  {{\em J. Phys. Conf. Ser.} \textbf{2015},
  {\em 626}, 012041}.

\bibitem{Mota:2006fz}
Mota, D.F.; Shaw, D.J.  {Evading equivalence principle violations,
  astrophysical and cosmological constraints in scalar field theories with a
  strong coupling to matter}.
{{\em Phys. Rev. D} \textbf{2007}, {\em
  75}, 063501}.

\bibitem{Waterhouse:2006wv}
Waterhouse, T.P.    {\em An Introduction to Chameleon Gravity}; The Cornell University:  Ithaca, NY, \mbox{USA, 2006.}


\bibitem{Weltman:2008ll}
Weltman, A.   {\em Studies in String Cosmology}; VDM Publishing: {Saarbr\"{u}cken, Germany},~2008.

\bibitem{Upadhye:2012fz}
Upadhye, A.   {\em Particles and Forces from Chameleon Dark Energy}; The Cornell University:  Ithaca, NY, USA, 2012.

\bibitem{Roy:2015cna}
Roy, N.; Banerjee, N.
   {Dynamical systems study of chameleon scalar field}.
  {{\em Ann. Phys.} \textbf{2015},
  {\em 356},~452--466}.

\bibitem{Mota:2006ed}
Mota, D.F.; Shaw, D.J. {Strongly coupled chameleon fields: New horizons
  in scalar field theory}.
  {{\em Phys. Rev. Lett.} \textbf{2006},
  {\em 97},  151102}.

\bibitem{Mota:2003tc}
Mota, D.F.; Barrow, J.D. {Varying alpha in a more realistic universe}.
 {\em Phys. Lett. B} \textbf{2004},
  {\em 581},  141--146.

\bibitem{Mota:2003tm}
Mota, D.F.; Barrow, J.D.  {Local and Global Variations of The Fine
  Structure Constant}.
 {{\em Mon. Not. R.
  Astron. Soc.}  \textbf{2004}, {\em 349},  291--302}.

\bibitem{Nelson:2008tn}
Nelson, A.E.; Walsh, J. {Chameleon Vector Bosons}.
  {{\em Phys. Rev. D} \textbf{2008}, {\em
  77},  095006}. 

\bibitem{Copeland:2006wr}
Copeland, E.J.; Sami, M.; Tsujikawa, S. {Dynamics of dark energy}.
 {{\em Int. J. Mod. Phys. D} \textbf{2006},
  {\em 15},  1753--1936},

\bibitem{Brax:2004px}
Brax, P.; van de Bruck, C.; Davis, A.C.; Khoury, J.; Weltman, A.  {Chameleon
  dark energy}.
  {{\em \mbox{AIP Conf.
  Proc.}} \textbf{2005}, {\em 736},  105--110}.

\bibitem{Brax:2004qh}
Brax, P.; van de Bruck, C.; Davis, A.-C.; Khoury, J.; Weltman, A. {Detecting
  dark energy in orbit: The cosmological chameleon}.
  {{\em Phys. Rev. D} \textbf{2004}, {\em
  70},  123518}.

\bibitem{Gubser:2004uf}
Gubser, S.S.; Khoury, J. {Scalar self-interactions loosen constraints from
  fifth force searches}.
  {{\em \mbox{Phys. Rev. D}} \textbf{2004}, {\em
  70}, 104001}.

\bibitem{Capozziello:2011et}
Capozziello, S.; de Laurentis, M. {Extended Theories of Gravity}.
 {{\em Phys. Rep.} \textbf{2011}, {\em
  509},  167--321}.

\bibitem{Capozziello:2007eu}
Capozziello, S.; Tsujikawa, S.
  {Solar system and equivalence principle constraints on f(R) gravity by chameleon approach}.
{{\em Phys. Rev. D} \textbf{2008}, {\em
  77},  107501}.

\bibitem{Faul}
Faulkner, T.; Tegmark, M.; Bunn, E.F.; Mao, Y.
{Constraining f(R) gravity as a scalar tensor theory}.
{{\em Phys. Rev. D} \textbf{2007}, {\em
  76},  063505}.

\newpage
\bibitem{Fujii}
Fujii, Y. \textit{The Scalar-Tensor theory of gravitation};  Cambridge University Press: Cambridge, UK,~2003.

\end{thebibliography}
\makeatletter
\renewcommand\@biblabel[1]{#1. }
\makeatother

\end{document}